\documentclass[preprint,12pt,authoryear]{elsarticle}

\usepackage{latexsym}
\usepackage{amssymb}
\usepackage{amsmath}
\usepackage{amsthm}
\usepackage{booktabs}
\usepackage{enumitem}
\usepackage{graphicx}
\usepackage{xcolor}
\usepackage{listings}
\usepackage{dsfont}
\usepackage{subcaption}
\usepackage{float}
\usepackage{dblfloatfix}

\usepackage{multirow}
\usepackage[normalem]{ulem}

\newcommand{\inc}[1]{\uwave{\emph{#1}}}

\usepackage{siunitx}
\usepackage[hidelinks]{hyperref}
\usepackage{cleveref}

\biboptions{authoryear,round}

\begin{document}
\begin{frontmatter}
    \title{Transition-Related Potentials as Markers of\\Narrative Comprehension in Continuous EEG}

    \author[elte]{Bálint Csanády\corref{cor1}}
    \author[bme,wigner]{Péter Vedres}
    \author[elte]{Kristóf Zsolt Makó}
    \author[elte,semmelweis]{\\Orsolya Papp-Zipernovszky}
    \author[szeged]{Márta Volosin}
    \author[elte]{Dávid Apagyi}
    \author[elte]{\\András Lukács}
    \author[elte]{András Bálint Kovács}
    \author[elte,miskolc,utexas]{Zoltan Nadasdy\corref{cor2}}
    
    \cortext[cor1]{Corresponding author: csbalint@protonmail.ch}
    \cortext[cor2]{Corresponding author: zoltan@utexas.edu}

    \affiliation[elte]{
        organization={ELTE Eötvös Loránd University},
        city={Budapest},
        country={Hungary}
    }

    \affiliation[bme]{
        organization={Budapest University of Technology and Economics},
        city={Budapest},
        country={Hungary}
    }

    \affiliation[wigner]{
        organization={HUN-REN Wigner Research Centre for Physics},
        city={Budapest},
        country={Hungary}
    }

    \affiliation[semmelweis]{
        organization={Semmelweis University},
        city={Budapest},
        country={Hungary}
    } 

    \affiliation[miskolc]{
        organization={University of Miskolc},
        city={Miskolc},
        country={Hungary}
    }

    \affiliation[szeged]{
        organization={University of Szeged},
        city={Szeged},
        country={Hungary}
    }
    
    \affiliation[utexas]{
        organization={The University of Texas at Austin},
        city={Austin},
        state={TX},
        country={USA}
    }    
           
    \begin{abstract}
        Harnessing the potential of electroencephalography (EEG) for brain research is fundamentally limited by intrinsic noise and the diffuse projection of brain-generated activity over the scalp.
        The standard event-related potential (ERP) paradigm addresses this limitation by relying on repeated independent trials, albeit at the cost of moving away from naturalistic experimental conditions.
        As a more naturalistic alternative, we collected continuous EEG while participants watched short films and extracted potentials aligned to sharp cinematic transitions (cuts).
        We demonstrate that such transition-related potentials (TRPs) exhibit canonical ERP-like temporal structure associated with significant information processing.
        By comparing coherent films with scene-scrambled versions containing matched post-cut sensory input, we find that these responses are systematically shaped by narrative context.
        We then show that the cut-related EEG signature can be recovered directly from group-averaged continuous recordings with a compact deep neural network (DNN).
        The detector generalized across films and subject groups, and the resulting TRPs reproduced the main context-dependent effects observed for manually annotated cuts.
        These results indicate that narrative context leaves a measurable signature in EEG responses,
        that this signature can be detected directly in continuous recordings, and that such detections provide a semi-automated framework for analyzing how viewers process and understand film narratives.
        We propose that the method outlined here can be adapted to parse EEG responses to other forms of continuous stimulation, providing a general tool for probing experimental conditions that are closer to natural human experience.
    \end{abstract}

\end{frontmatter}


\section{Introduction}
    Our conscious experience is dominated by the parsing of continuous sensory streams into discrete, meaningful segments, through which we construct the underlying narrative structure of the world around us.
This segmentation is often triggered by sensory events that elicit recursive perceptual cycles of feature extraction,
scene segmentation, object recognition, and integration with prior knowledge \citep{Nadasdy2024Event}.
Understanding the neuronal processes underlying such events remains a challenge in cognitive neuroscience, especially when relying on non-invasive and readily available tools such as electroencephalography (EEG).
What triggers a cognitive event in one individual may not trigger the same event in another, and even when an event is shared, the timing of the corresponding cognitive EEG markers may differ across people.
This variability is compounded by the fact that scalp-recorded activity has a low signal-to-noise ratio (SNR) and reflects the simultaneous activation, interference, and superposition of multiple neural sources.
Moreover, the mapping from neural sources to scalp potentials is not uniquely invertible,
so recovering the underlying neural events from continuous EEG constitutes an ill-posed inverse problem \citep{nunez2006electric}.

A standard way to separate stimulus-related activity from noise is to improve the SNR by averaging EEG segments aligned to repeated triggers.
In the event-related potential (ERP) paradigm, this trigger is usually a stimulus onset.
EEG segments aligned to repeated presentations of the same stimulus or stimulus category allow related activity to combine constructively,
while activity unrelated to the stimulus is attenuated \citep{luck2014introduction,kappenman2016best,boudewyn2018many}.
This yields reproducible ERP waveforms and has become a central method for tracking cognitive processes with EEG.
However, the method typically relies on a trial-by-trial structure in which stimuli are repeatedly presented in isolation, usually in a randomized order, so that trials can be treated as approximately independent.
This requirement is in tension with cognition in its natural form, where each event unfolds in relation to those preceding it.
In this way, the continuous experience that motivates the question is partly sacrificed for methodological control.

In contrast, allowing subjects to watch short films as a continuous, task-free visual experience provides a more naturalistic window into brain activity \citep{SONKUSARE2019699}.
Although professionally produced films differ markedly from the trial-by-trial stimulus presentations typical of ERP studies, they are far from unstructured stimuli.
Films contain precisely planned transitions (cuts), and salient yet unpredictable events that closely emulate real-life scenarios.

We recorded EEG from participants while they watched a film, treated the film as a continuous naturalistic stimulus, and used the cuts separating shots as event triggers.
The method preserves the continuity and contextual dependence of naturalistic experience, but introduces transitions that are discrete, frame-accurate, and shared across viewers.
While individual differences in interpretation may introduce substantial inter-individual EEG variance \citep{mars2008trial},
cuts serve as precisely timed markers around which average waveforms can be computed,
approximating the kinds of event boundaries that are thought to structure ongoing experience \citep{zacks2007eventperception,zacks2010cuttingroom,kurby2008segmentation,magliano2011continuityediting}.
Related event-boundary and cinematic-transition studies have shown that such transitions can elicit characteristic ERP-like responses \citep{silva2019rapid,sanz2023cinematographic}.
As such, cuts provide temporal anchors for EEG analysis through transition-related potentials (TRPs), marking moments at which attention may shift and new information may be processed.

If cuts can serve as temporal anchors for naturalistic EEG, the next question is what kind of neural response they elicit.
The EEG responses to cuts themselves have been characterized in some detail.
Cut-locked responses contain both early components 
and later components 
\citep{matran2015event}; they are sensitive to whether a cut joins related or unrelated material \citep{francuz2013does}, and they vary with the editing technique used \citep{heimann2017cuts}.
What has not yet been examined is whether and how these responses depend on narrative coherence, that is, on the extent to which transitions are supported by the preceding narrative context.
Furthermore, to date, these responses have primarily been measured by averaging around cuts whose timing was already known.
It remained an open question whether cut-related responses can be detected within the continuous EEG recordings without prior knowledge of the cut timings.

In this study, we investigate both questions.
First, we characterize the effect of narrative coherence on cut-locked TRP signals.
Second, we propose a population-level deep-learning model for detecting cut-related EEG responses in continuous recordings.
In addition to the model's ability to identify cuts with great precision,
we demonstrate that the detected timestamps are suitable substitutes for the original cut annotations,
in the sense that the resulting TRPs still exhibit the same characteristics in relation to narrative coherence as their hand-annotated counterparts.
Finally, we compare our cut-detection approach with alternative methods to better understand the factors driving model performance.
Together, these analyses provide a semi-automated framework for using EEG to study how viewers segment, update, and comprehend film narratives during naturalistic viewing. 

\section{Experimental Setup}
    The experiment was conducted at the Institute of Psychology, University of Szeged, between October 2022 and May 2023.
Participants sat in a comfortable chair in a dimly lit, sound-attenuated, and electrically shielded room.
A 20-inch LCD monitor (LG Flatron L206WTQ-SF; LG Electronics) and two speakers were placed on a table in front of them.
Participants were presented with one of two short romantic films with minimal dialogue and a clear narrative structure: \emph{The Art of Love} \citep{farinella2013artoflove} or \textit{City Lights} \citep{wiles2016citylights}.
Hereafter, we refer to them as \emph{Art} and \emph{City}, respectively.

\begin{figure*}[ht]
     \centering
     \includegraphics[width=\textwidth]{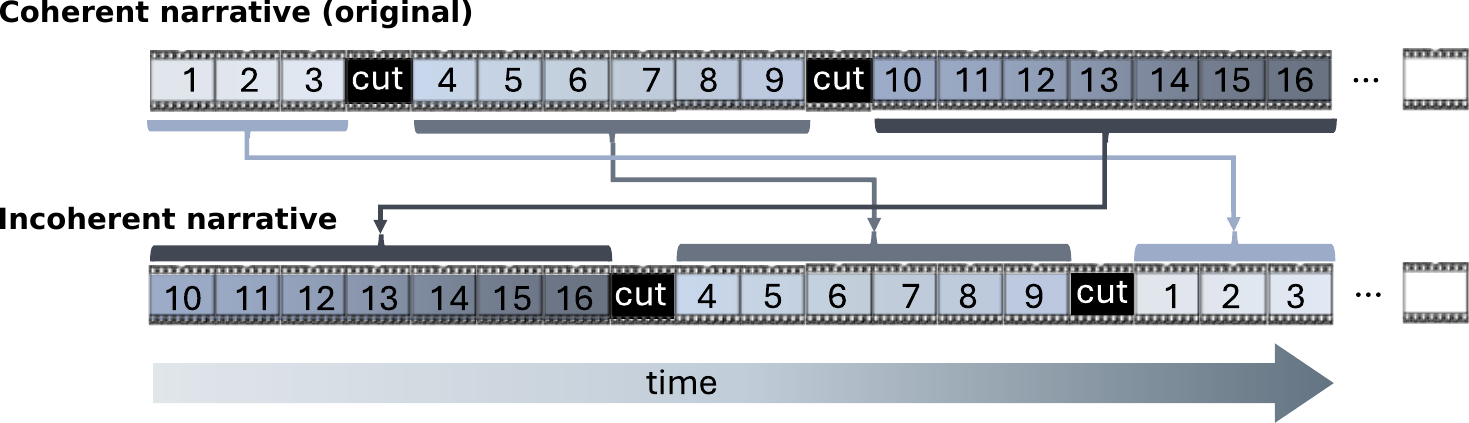}
     \caption{Editing the films with incoherent narratives. The incoherent versions were created by shuffling the takes separated by cuts.
     Each participant watched one version of each film, one with a coherent narrative and one with an incoherent narrative.
     }
    \label{fig:cutting}
\end{figure*}

In addition to the original film with a coherent narrative, an incoherent version was created by a professional editor, who cut the film at scene boundaries and randomly reordered the scenes.
Because simple randomization could leave two consecutive takes in place, we enforced a scrambling procedure that preserved no consecutive takes (Figure~\ref{fig:cutting}).
Each participant viewed the coherent version of one film and the incoherent version of the other.
Thus, participants watched either the coherent version of \emph{Art} and the incoherent version of \emph{City}, or vice versa.
The order of film presentation (film title and narrative coherence) was counterbalanced across participants.
After each film, participants completed the validated Hungarian version of the Narrative Engagement Scale \citep{papp2024narrativ}.
The Narrative Engagement Scale is a 12-item self-report measure of audience engagement with narrative content, originally developed by \citet{busselle2009measuring}.
It contains four subscales (Narrative Understanding, Attentional Focus, Emotional Engagement, and Narrative Presence), and each response is given on a 7-point Likert scale.

\subsection{Participants}
    The experiment included $34$ university students aged between 18 and 47 years (10 men, 24 women; mean age: $22.82$ years, SD: $5.48$; $5$ left-handed).
    Participants were recruited through university courses and mailing lists, and received no financial compensation.
    All reported normal or corrected-to-normal vision, normal hearing, and no history of psychiatric or neurological disorders.
    None were studying cinematography or related fields.
    All participants provided written informed consent after receiving a full explanation of the experimental procedure.
    The study was approved by the United Review Committee for Research in Psychology (Hungary, EPKEB – number 2021-123), and was conducted in accordance with the Declaration of Helsinki \citep{WMA2025DeclarationHelsinki}.

\subsection{EEG Recording}
    Continuous EEG was recorded at a sampling rate of $512\,\mathrm{Hz}$ using a BioSemi ActiveTwo system \citep{BioSemiActiveTwo}.
    A total of 34 active electrodes (32 scalp electrodes and 2 additional electrodes) were mounted on an elastic cap according to the 10/20 system \citep{nuwer1998ifcn}.
    Electrode impedances were kept below 30 k$\Omega$.
    Stimulus presentation was controlled using MATLAB (Version 9.0.0, R2016a; The MathWorks Inc.) and Psychophysics Toolbox Version 3.0.18 \citep{brainard1997psychophysics,kleiner2007s}, running under the Windows 10 operating system.

\subsection{EEG Preprocessing}
    The EEG data were processed in MNE-Python \citep{gramfort2013meg} using the BioSemi 32-channel montage.
    Cuts were manually annotated at frame accuracy and stored as millisecond timestamps corresponding to the start of the first frame following each cut.
    Bad channels were automatically identified in MNE on the basis of a local outlier factor criterion and subsequently reconstructed by spatial interpolation.
    Line noise was removed by a $50\,\mathrm{Hz}$ notch filter, and a $0.1$--$60\,\mathrm{Hz}$ band-pass filter was applied to reduce low-frequency drift and high-frequency noise.
    Extracted EEG segments were normalized per channel by dividing by the standard deviation computed over the full epoch window ($-200$ to $1200\,\mathrm{ms}$), then baseline-corrected by subtracting the mean amplitude of the $-200$ to $0\,\mathrm{ms}$ pre-stimulus interval.
    
\begin{figure*}[h!]
         \centering
         \begin{subfigure}[b]{.49\textwidth}
            \centering
            \includegraphics[width=\textwidth]{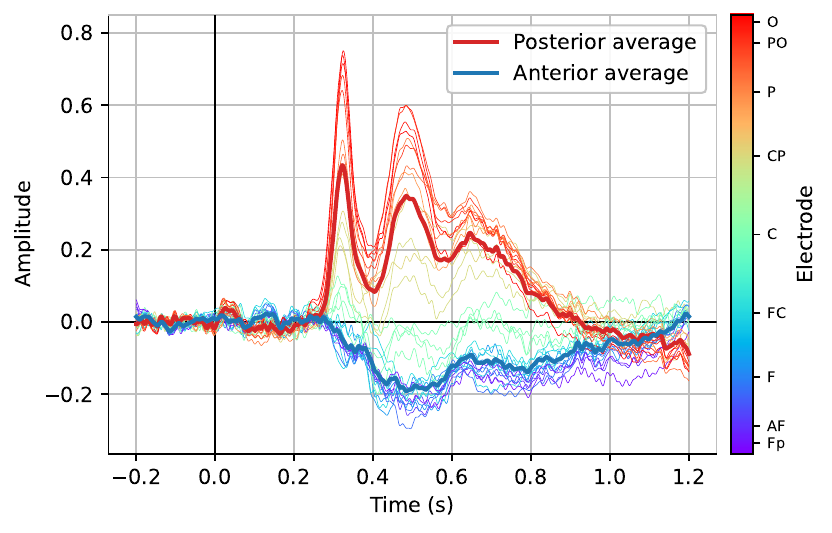}
            \subcaption{\emph{Art}}
            \label{fig:Butterfly_Art_Coherent}
         \end{subfigure}
         \hfill
         \begin{subfigure}[b]{.49\textwidth}
            \centering
            \includegraphics[width=\textwidth]{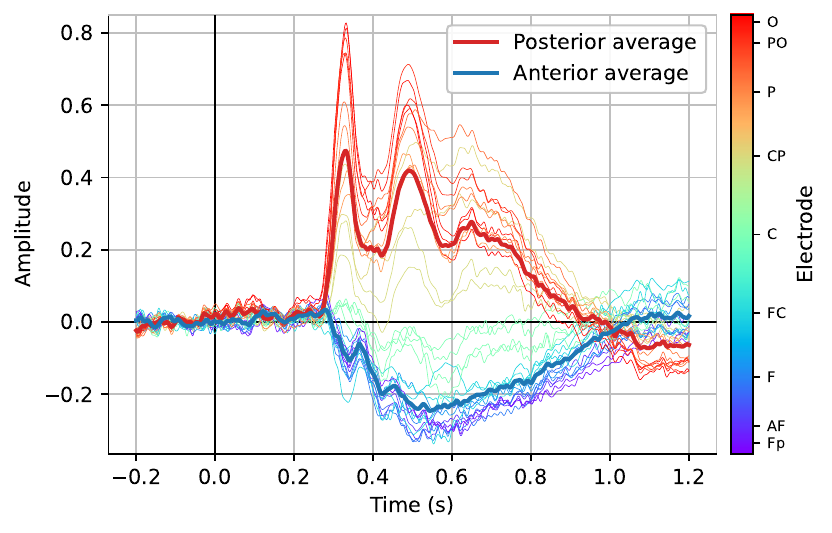}
            \subcaption{\emph{City}}
            \label{fig:butterfly_City_Gold_ap}
         \end{subfigure}
         \caption{Butterfly plots of the average TRP signals aligned to cuts in the coherent versions of the films, by individual electrodes and anterior/posterior averages.
         The color of the electrode traces is based on their anteriority.
         }
        \label{fig:Gold_AP}
    \end{figure*}

    To visualize the spatial distribution of the cut-TRP response, electrode-wise grand-average waveforms were computed separately for each film by averaging across participants.
    Figure~\ref{fig:Gold_AP} shows the resulting butterfly plots for the coherent versions of the films, together with the corresponding anterior and posterior averages.
    A separation between the anterior (negative) and posterior (positive) electrode groups is clearly visible, motivating the use of anterior and posterior electrode-group averages in the TRP analyses below.
    
\section{Results}
    \subsection{TRP Analysis}\label{sec:TRP}
        Rather than partitioning the experiment into discrete trials, we annotated the timing of the cuts in the films and treated them as naturally occurring temporal markers in the continuous EEG recordings.
Because cinematic cuts can take several forms, ranging from gradual dimming or translucent transitions to instantaneous frame changes without an intervening blank frame, we defined each cut as the earliest detectable transition between two scenes.
If cuts function as event boundaries at which significant new information is introduced, the corresponding cut-locked TRPs should contain ERP-like components, interpreted within the standard ERP framework \citep{luck2014introduction,picton2000erp_guidelines}.

As anticipated, the TRP displayed in Figure~\ref{fig:Gold_AP} reproduced a prototypical posterior positivity in the P2 range ($\sim\qtyrange{250}{400}{\milli\second}$), followed by prominent late components including the P3/P300 ($\sim\qtyrange{400}{600}{\milli\second}$) and LPC/P600 ($\gtrsim\qty{600}{\milli\second}$) over the posterior domain.
This was complemented over the anterior electrodes by a modest N270-like deflection and a sustained late negativity between $400$ and $\qty{1000}{\milli\second}$.
One striking feature of the TRPs was a divergence between anterior and posterior activity.
This divergence began with a steep increase in occipital positivity at approximately $\qty{300}{\milli\second}$ after cut onset,
leading to the peak formation of P2, and was accompanied by the simultaneous evolution of anterior negativity reaching the minimum at $\sim\qty{500}{\milli\second}$, followed by a convergence complete at $\sim\qty{1}{\second}$.
Conspicuously, the earlier exogenous visual components, such as P50 and N100, were not expressed.
This pattern is consistent with previous ERP studies using dynamic video stimuli, in which early visual components were weak or absent under continuous stimulus presentation, whereas later components related to scene comprehension, semantic integration, and post-cut updating remained robustly detectable \citep{sitnikova2003semantic,sitnikova2008two,sanz2023cinematographic}.
These early sensory components are likely attenuated, refractory, or temporally smeared in continuous film viewing, because the visual system is already engaged by ongoing stimulation and because the exact perceptual registration of a cut may vary across edits and observers.
Despite the possible arbitrariness of onset detection, the ERP-like components were reproduced with temporal precision characteristic of event-related responses.

\begin{figure*}[h!]
     \centering
     \begin{subfigure}[b]{.49\textwidth}
        \centering
        \includegraphics[width=\textwidth]{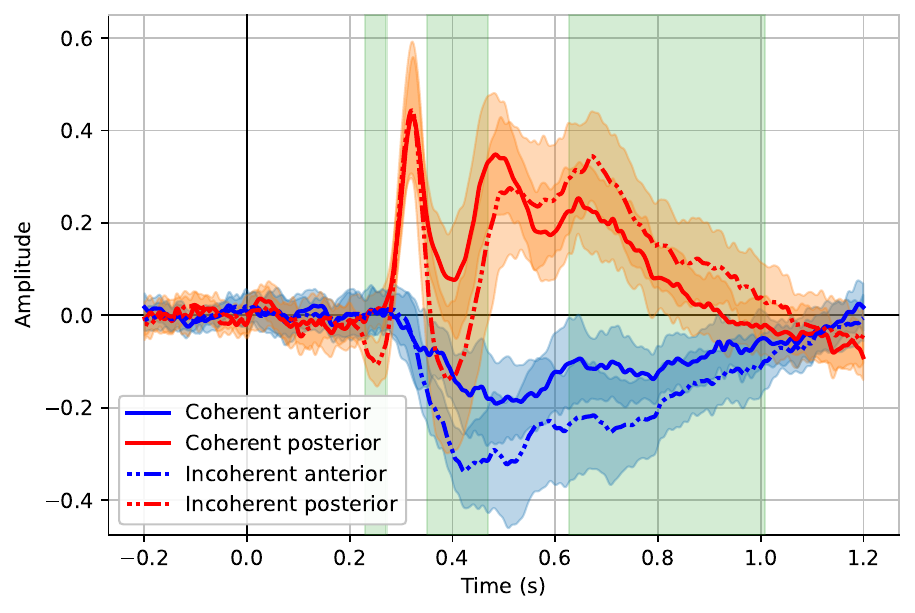}
        \subcaption{\emph{Art}}
        \label{fig:Clutsering_Art_l_v_nl}
     \end{subfigure}
     \hfill
     \begin{subfigure}[b]{.49\textwidth}
        \centering
        \includegraphics[width=\textwidth]{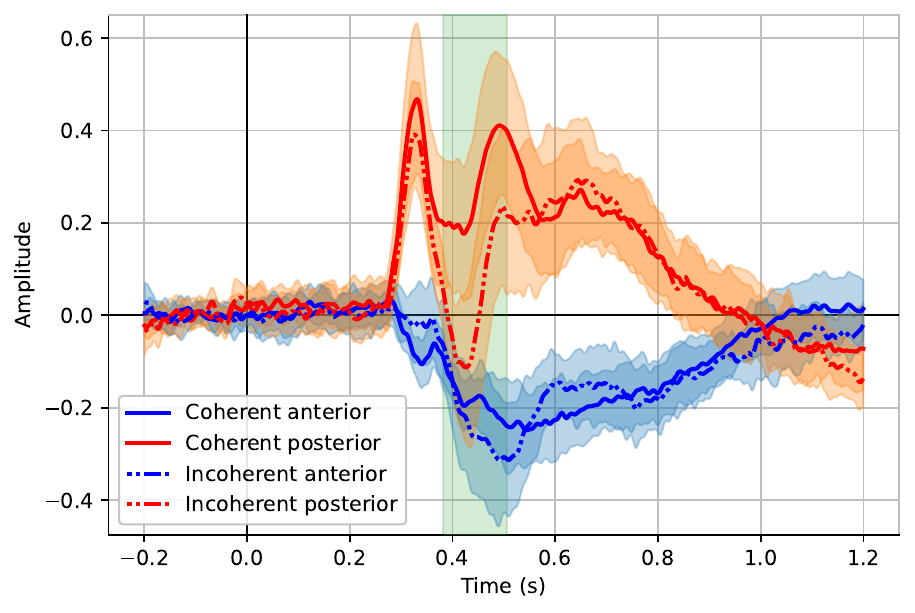}
        \subcaption{\emph{City}}
        \label{fig:Clutsering_City_l_v_nl}
     \end{subfigure}
     \caption{Anterior and posterior TRPs of the cuts within coherent and incoherent narrative versions of the same film. Cuts are matched one-to-one. Green bars indicate significant clusters.}
    \label{fig:Clutsering_Gold}
\end{figure*}

To investigate whether and how the cut-locked response was shaped by the coherence of the film's narrative, we compared TRPs derived from the coherent and incoherent versions of each film.
Since the cut-locked TRPs demonstrate a clear separation of anterior and posterior activity, we compared the anterior and posterior average TRP signals (Figure~\ref{fig:Clutsering_Gold}).
Clear statistical differences were observed between TRP amplitudes following coherent vs. incoherent cuts, highlighted by green shaded intervals.

Statistical significance was assessed using a spatio-temporal cluster permutation test across subjects \citep{maris2007nonparametric}.
Candidate clusters were formed across adjacent time samples and neighboring EEG sensors using a one-way F-statistic with a cluster-forming threshold of $p < 0.0027$ ($3\sigma$).
Clusters were considered significant at a threshold of $p < 0.0456$ ($2\sigma$), estimated using $\num{10000}$ permutations.
This conservative cluster-forming threshold was chosen to favor focal, high-confidence effects over broad clusters driven by weaker distributed differences.
In both films, the posterior average showed a stronger late negative deflection when the narrative was incoherent.
Additionally, in \emph{Art}, there was a statistically significant negative bump in the incoherent posterior average, and the amplitude attenuated more slowly during cuts related to an incoherent narrative context.
The exact details of the clusters are shown in Table~\ref{tab:art_city_clusters}.

\begin{table}[h!]
    \centering
    \resizebox{\textwidth}{!}{
    \begin{tabular}{l|cccc}
    \noalign{\hrule height 1pt}
    Film & $t_{\min}$ & $t_{\max}$ & $p$-value & Channels \\
    \noalign{\hrule height 1pt}
    \emph{Art}  & 0.230 & 0.271 & 0.0106 & \footnotesize{\texttt{CP1,CP2,CP5,CP6,O1,O2,Oz,P3,P4,P7,P8,PO3,PO4,Pz}}\\
    \emph{Art}  & 0.350 & 0.469 & 0.0018 & \footnotesize{\texttt{AF3,AF4,C4,CP5,CP6,F3,F7,F8,FC5,FC6,Fp1,Fp2,P4,P7,T7,T8}}\\
    \emph{Art}  & 0.627 & 1.008 & 0.0011 & \footnotesize{\texttt{CP1,CP2,Cz,F3,F7,FC1,FC5,P3,P4,PO3,Pz,T7}}\\
    \hline
    \emph{City} & 0.381 & 0.506 & 0.0042 & \footnotesize{\texttt{CP1,CP2,CP5,CP6,F8,FC6,O1,Oz,P3,P4,P7,P8,PO3,PO4,Pz,T7,T8}}\\
    \noalign{\hrule height 1pt}
    \end{tabular}
    }
    \caption{Significant clusters for the \emph{Art} and \emph{City} films.}
    \label{tab:art_city_clusters}
\end{table}

It is worth pointing out that in these comparisons the film segments following the cuts were paired one-to-one with frame precision.
This means that the TRPs compared aggregate responses from time windows in which the immediate post-cut audiovisual information presented to participants was identical across the coherent and incoherent versions.
As noted above, coherence assignment was counterbalanced across films, so the observed effects cannot be attributed to a fixed difference between participant groups.
Consequently, the measured effects reflect the influence of preceding context, including narrative comprehension.

To assess whether the differences observed over posterior and anterior electrode groups were related to participants’ engagement with the narrative,
we analyzed the post-experiment questionnaire completed by each participant.
Table~\ref{tab:narrative_engagement} shows the aggregate Narrative Engagement scores from the questionnaire completed after each film.
At the individual level, coherent film viewing was strongly correlated with higher engagement scores (\emph{Art}: $0.77$; \emph{City}: $0.73$).
Overall, the coherent versions of the two films received significantly higher engagement scores than their incoherent versions, providing a plausible behavioral correlate of the TRP differences.
Moreover, people who watched the incoherent version of \emph{Art} scored somewhat lower on the scale, which is consistent with the stronger TRP difference; however, the confidence intervals overlap, so this effect cannot be taken as conclusive.

\begin{table}[h!]
    \centering
    \begin{tabular}{l|cc|cc}
        \noalign{\hrule height 1pt}
        \textbf{Narrative} & \multicolumn{2}{c|}{\textbf{Coherent}} & \multicolumn{2}{c}{\textbf{Incoherent}} \\
        \textbf{Film} & \textbf{\emph{Art}} & \textbf{\emph{City}} & \textbf{\emph{Art}} & \textbf{\emph{City}} \\
        \noalign{\hrule height 1pt}
        \textbf{Mean score}              & 64.2 & 66.3 & 42.6 & 45.3 \\
        \textbf{$\mathrm{CI}_{95\%}$}    & 4.8  & 3.1  & 4.5  & 7.5  \\
        \noalign{\hrule height 1pt}
    \end{tabular}
    \caption{Narrative Engagement mean scores and confidence intervals by film.}
    \label{tab:narrative_engagement}
\end{table}

\begin{figure*}[h!]
     \centering
     \begin{subfigure}[b]{.48\textwidth}
        \centering
        \includegraphics[width=\textwidth]{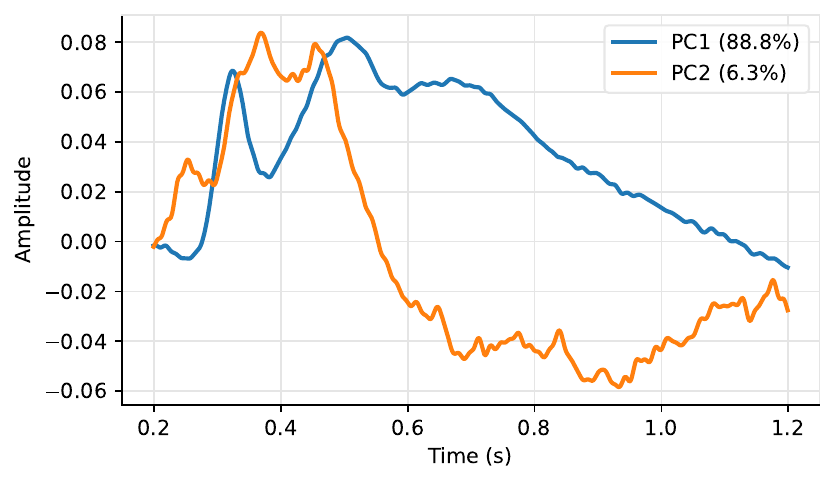}
        \subcaption{\emph{Art} principal eigenvectors.}
        \label{fig:Art_eigenvectors}
     \end{subfigure}
     \hfill
     \begin{subfigure}[b]{.48\textwidth}
        \centering
        \includegraphics[width=\textwidth]{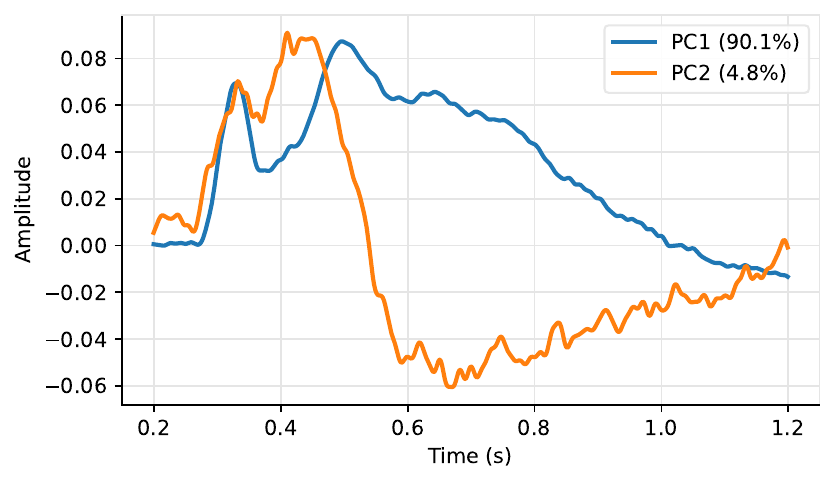}
        \subcaption{\emph{City} principal eigenvectors.}
        \label{fig:City_eigenvectors}
     \end{subfigure}
     \vskip 3mm
     \begin{subfigure}[b]{.24\textwidth}
        \centering
        \includegraphics[width=\textwidth]{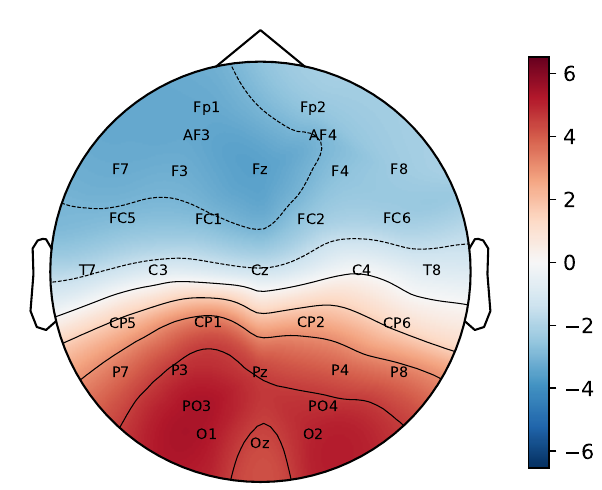}
        \subcaption{\emph{Art} coherent PC1.}
     \end{subfigure}
     \hfill
     \begin{subfigure}[b]{.24\textwidth}
        \centering
        \includegraphics[width=\textwidth]{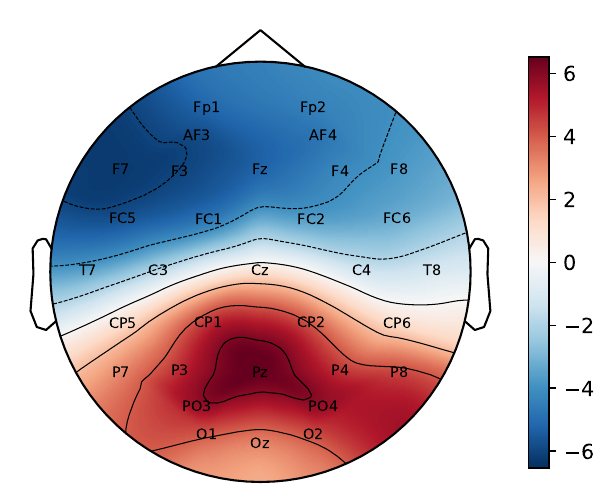}
        \subcaption{\emph{Art} incoherent PC1.}
     \end{subfigure}
     \hfill
     \begin{subfigure}[b]{.24\textwidth}
        \centering
        \includegraphics[width=\textwidth]{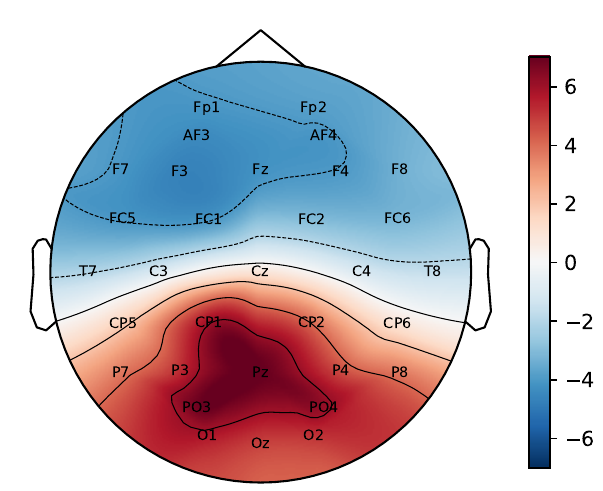}
        \subcaption{\emph{City} coherent PC1.}
     \end{subfigure}
     \hfill
     \begin{subfigure}[b]{.24\textwidth}
        \centering
        \includegraphics[width=\textwidth]{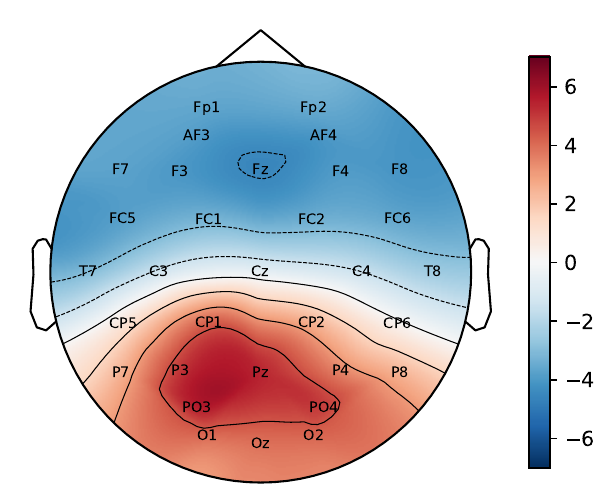}
        \subcaption{\emph{City} incoherent PC1.}
     \end{subfigure}
     \vskip 3mm
     \begin{subfigure}[b]{.24\textwidth}
        \centering
        \includegraphics[width=\textwidth]{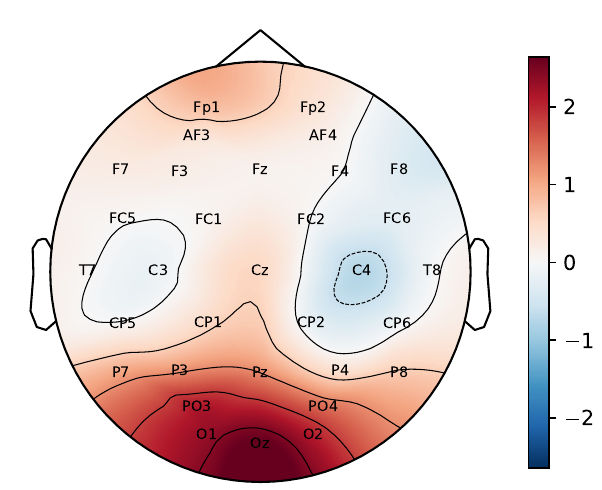}
        \subcaption{\emph{Art} coherent PC2.}
     \end{subfigure}
     \hfill
     \begin{subfigure}[b]{.24\textwidth}
        \centering
        \includegraphics[width=\textwidth]{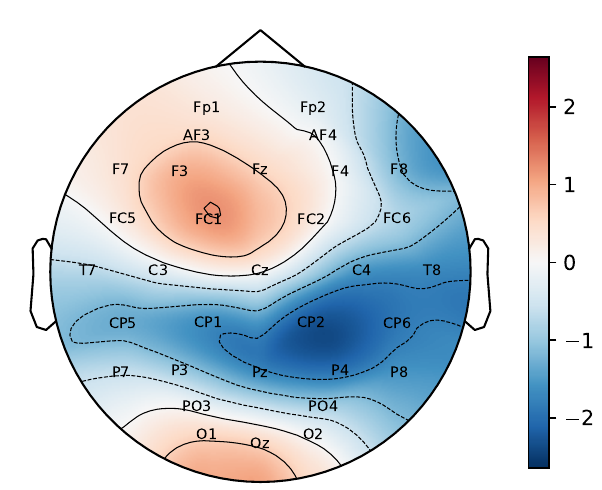}
        \subcaption{\emph{Art} incoherent PC2.}
     \end{subfigure}
     \hfill
     \begin{subfigure}[b]{.24\textwidth}
        \centering
        \includegraphics[width=\textwidth]{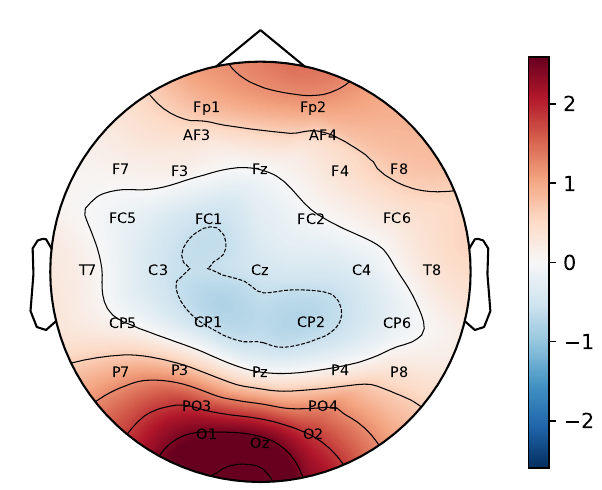}
        \subcaption{\emph{City} coherent PC2.}
     \end{subfigure}
     \hfill
     \begin{subfigure}[b]{.24\textwidth}
        \centering
        \includegraphics[width=\textwidth]{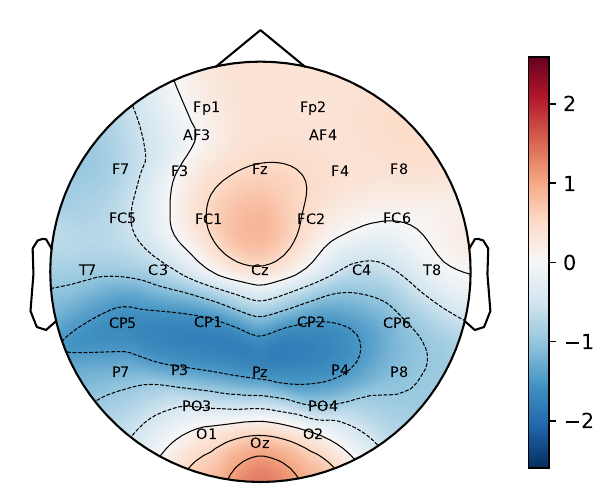}
        \subcaption{\emph{City} incoherent PC2.}
     \end{subfigure}
     \caption{Temporal PCA analysis of the TRPs.}
    \label{fig:PCA_temporal}
\end{figure*}

To elucidate the main factors underlying the modulation of TRP amplitudes, we performed a temporal Principal Component Analysis (PCA) on cut-locked TRP waveforms \citep{picton2000erp_guidelines}.
For a given film (e.g. \emph{Art}), we extracted the coherent and incoherent TRP channel averages from $\num{200}$ to $\qty{1200}{\milli\second}$.
This yielded 64 observations in a 512-dimensional temporal feature space.
PCA on these observations produced 512-dimensional temporal eigenvectors and principal component (PC) scores consisting of $32+32$ weights, one for each channel in each condition.
Figure~\ref{fig:PCA_temporal} summarizes the results of this analysis, performed on \emph{Art} and \emph{City} independently.
Notably, the resulting eigenvectors were similar across films, and the first PC scores highlight the main anterior--posterior polarity separation we have already remarked on. 
This evidence supports the anterior--posterior polarity separation as the strongest topographic pattern associated with the processing of cuts in both coherent and incoherent narratives.
The second PC, in contrast, captured a posterior topographic pattern related to narrative coherence (Figure~\ref{fig:PCA_temporal}).
We can see that PC2 modulates the extensive temporal-parietal negativity of the occipito-parieto-temporal dipole for the incoherent narratives for both films while it was completely absent during watching the coherent films.
A spatial variant of the PCA analysis is available in the Appendix (Figure~\ref{fig:PCA_Spatial}).

    \subsection{Cut Detection}\label{sec:cut_detect}
        We have shown that TRPs are sensitive markers of cognitive events that not only reproduce critical features of ERPs,
but also reveal important aspects of film narrative processing without participants being explicitly instructed to pay attention to them.
This finding motivated us to use deep learning to detect significant cognitive events that occur naturally during the processing of the complex sensory stream typical of film viewing.
More specifically, we trained a deep neural network (DNN) to detect manually annotated cinematic cuts from EEG signals alone and then applied the model to recover cut-like EEG patterns,
including patterns not necessarily elicited by cinematic cuts.

DNNs are well established for detecting events in EEG, most prominently for seizure detection \citep{roy2019deep} and sleep-stage scoring \citep{supratak2017deepsleepnet},
and compact architectures developed for brain-computer interfaces have shown that such models can generalize across tasks and subjects \citep{lawhern2018eegnet, schirrmeister2017deep}.
These events, however, are defined by the neural state itself: seizures and sleep stages can be identified from EEG by professionals, which guarantees the presence of a detectable neural signature.
In contrast, cinematic cuts are defined objectively by the film and independently of brain activity.
In addition, cuts are audiovisual transitions that may elicit sensory and cognitive processes related to the closure of the previous event and the beginning of a new one,
consistent with event-segmentation accounts of film comprehension and cinematic continuity \citep{zacks2007eventperception,zacks2010cuttingroom,magliano2011continuityediting,smith2012attentionalcontinuity},
thereby modulating late ERP-like components.
Studies using naturalistic stimuli have applied inter-subject correlation to identify stimulus-driven EEG responses shared across viewers \citep{hasson2004intersubject}.
Our analysis also relied on common components of EEG signals, but focused on a more complex objective.
We asked not only whether the events could be detected with accuracy exceeding correlation-based methods, but also whether their onsets could be localized with sufficient temporal precision to recover high-quality TRPs.

\begin{figure*}[h!]
     \centering
     \begin{subfigure}[b]{.98\textwidth}
        \centering
        \includegraphics[width=\textwidth]{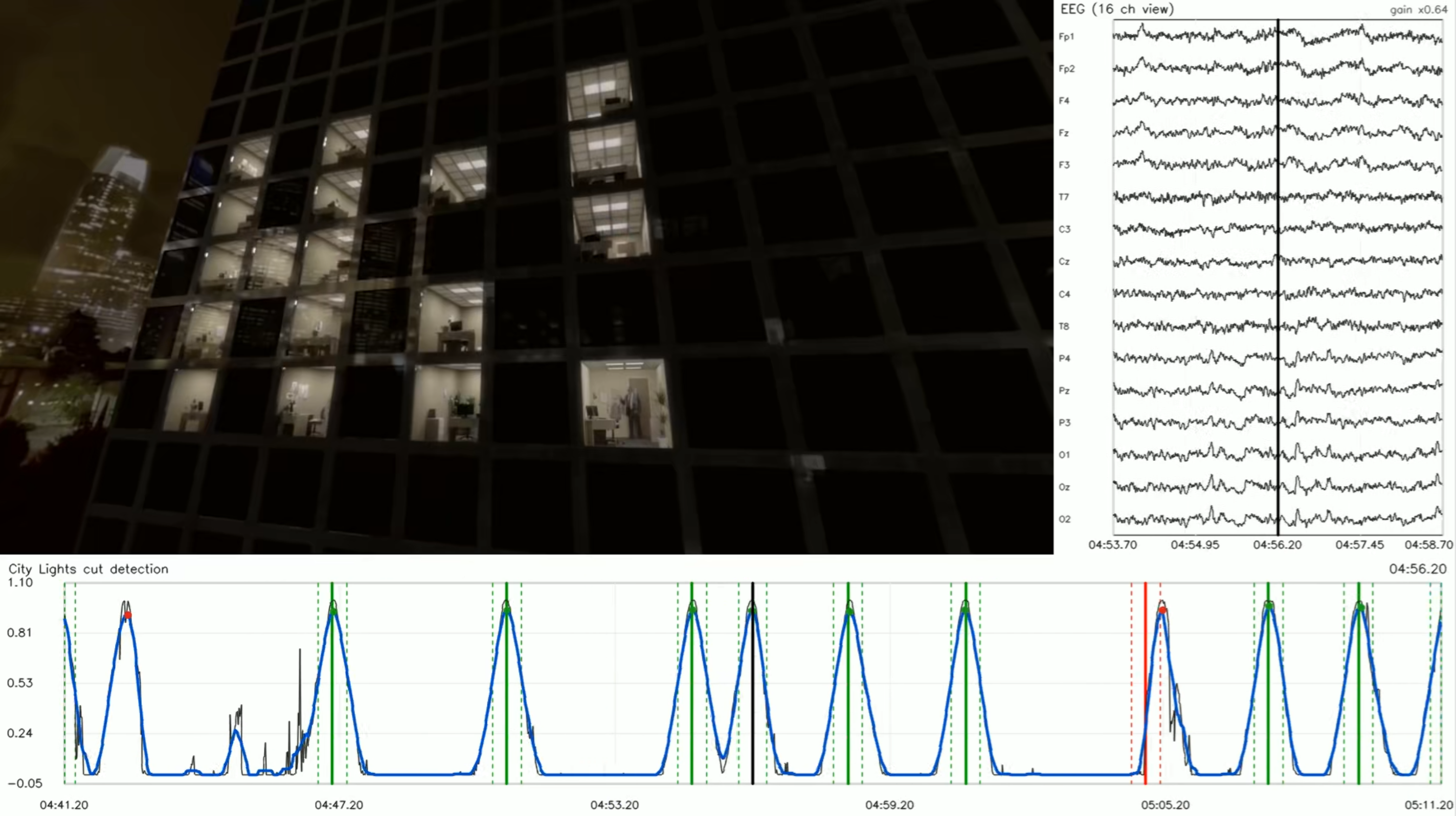}
     \end{subfigure}
     \caption{Snapshot from the EEG cut detection on the film \emph{City}.}
    \label{fig:cut_detection}
\end{figure*}

To fulfill these requirements, we trained a DNN to recover the timing of cinematic cuts directly from the synchronized EEG signals.
The population-level detector operated on short sliding windows of the averaged EEG signal, producing a continuous cut-likelihood score that was converted into discrete cut predictions by detecting peaks in the score trace (Figure~\ref{fig:cut_detection}).
To clarify the detector, we first outline the DNN model used to compute the cut-likelihood score.

\subsubsection{DNN Model} 
    Let $X \in \mathbb{R}^{S \times 32 \times T}$ denote the cut-synchronized EEG recordings for one version of one film, where $S$ is the number of subjects, $32$ is the number of channels, and $T$ is the number of samples.
    To ensure comparable signal contribution from each participant, we first normalized each response using a causal rolling baseline Z-score at the participant level.
    Specifically, each EEG sample was standardized channel by channel relative to the mean and standard deviation of the preceding $\qty{200}{\milli\second}$ of the same signal.
    The normalized signals were then averaged across subjects to obtain $\hat X \in \mathbb{R}^{32 \times T}$.
    This across-subject averaging improved the signal-to-noise ratio by emphasizing activity that was temporally aligned and shared across subjects.
    $\hat X$ was divided into overlapping windows $W_i \in \mathbb{R}^{32 \times 512}$ of $512$ samples ($\qty{1}{\second}$).
   
    Each extracted window was assigned a ground-truth label $y(W_i)$ based on its temporal proximity to annotated events.
    For each window onset $t_i$ and nearest cut annotation $t^*_i$ (both measured in samples), we defined $y(W_i)$ as:

    $$
    y(W_i)=\min\left(\max\left(1.1-\frac{|t_i-t^*_i|}{320},0\right),1\right).
    $$

    \paragraph{Architecture}
        We built the cut detector as a compact classifier with recurrent convolutional architecture.
        A 1D convolutional embedding first compressed the input EEG window to a shorter temporal representation.
        We employed an embedding consisting of two convolutional stages with output channel sizes $32$ and $64$, using kernel size and stride of $3$ in the first and $4$ in the second layer.
        The embedded sequence was then processed by a two-layer bidirectional residual LSTM with $32$ hidden units.
        The processed sequence was average-pooled resulting in a single $64$ dimensional vector.
        Finally, this vector was processed by a dense classification head with $16$ hidden neurons, producing a scalar logit for each input window.
        The convolutional layers captured short-timescale temporal patterns, while the LSTM examined how these patterns evolved across the window.

    \paragraph{Training}
        The models were trained on randomly sampled mini-batches of size $64$, with binary cross-entropy loss on logits with the continuous target value $y(W_i)$, and AdamW optimization (learning rate: $0.0005$, weight decay: $0.001$).
        For sampling purposes, windows where $y(W_i)>0$ were considered positive, and windows with $y(W_i)=0$ were considered negative.
        To balance training, negative windows were randomly under-sampled in each epoch to match the number of positive windows.
        For regularization, we trained the networks using batch normalization and dropout ($0.25$). 
        To augment training, we added random white noise (SD: $0.1$) to each sample before presenting it to the network.
        Unless otherwise stated, models were trained for 20 epochs.

    \paragraph{Inference Methodology}
        At inference time, the model was evaluated at every eighth window, providing a time series of model predictions throughout the film.
        This model score trace was smoothed using a 20-sample moving average window.
        Cuts were predicted by peak finding on the smoothed score trace.
        A peak was accepted only if it satisfied the chosen minimum height (0.5), minimum width (3), and minimum inter-peak distance ($\qty{0.625}{\second}$).

        A detected cut was classified as a true positive (TP) if it fell within an acceptance window of $\qty{0.625}{\second}$ centered on a true cut.
        In other words a detected peak corresponding to EEG window $W_i$ was accepted if $y(W_i)\geq 0.575$.
        Similarly, detected cuts were classified as false positives (FPs) if they fell outside all acceptance windows, and acceptance windows with no detected peak were counted as false negatives (FNs).
        Double-counting of true positives was prevented by setting the minimum inter-peak distance equal to the acceptance-window width.
            
\subsubsection{Results}
    In the following analysis, we compared different ways of partitioning the four available datasets (2 films, 2 narrative coherence modes) into training and evaluation sets, as summarized in Table~\ref{tab:dnn_generalization_splits}.
    Each row defines one generalization setup.
    The DNN was trained on Group A and evaluated on Group B, and the same comparison was also performed in the reverse direction.
    These splits tested whether the detector generalized across film identity, participant group, narrative coherence, or the joint change of film identity and participant group.

    \begin{table}[h!]
        \centering
        \small
        \begin{tabular}{l|cc}
            \noalign{\hrule height 1pt}
            \textbf{Generalization} & \textbf{Group A} & \textbf{Group B} \\
            \noalign{\hrule height 1pt}
            Film &
            \emph{Art}, \inc{Art} &
            \emph{City}, \inc{City} \\
            Subjects &
            \emph{Art}, \inc{City} &
            \inc{Art}, \emph{City} \\
            Coherence &
            \emph{Art}, \emph{City} &
            \inc{Art}, \inc{City} \\
            \multirow{2}{*}{Film \& subjects} &
            \emph{Art} &
            \emph{City} \\
            &
            \inc{Art} &
            \inc{City} \\
            \noalign{\hrule height 1pt}
        \end{tabular}
        \caption{Dataset splits used for DNN generalization tests.
        Models were trained on one group and evaluated on the other, with both directions tested.
        The wavy underline marks the incoherent version of a film (e.g. \inc{Art}).}
        \label{tab:dnn_generalization_splits}
    \end{table}

    \begin{figure*}[h!]
        \centering
        \includegraphics[width=0.9\textwidth]{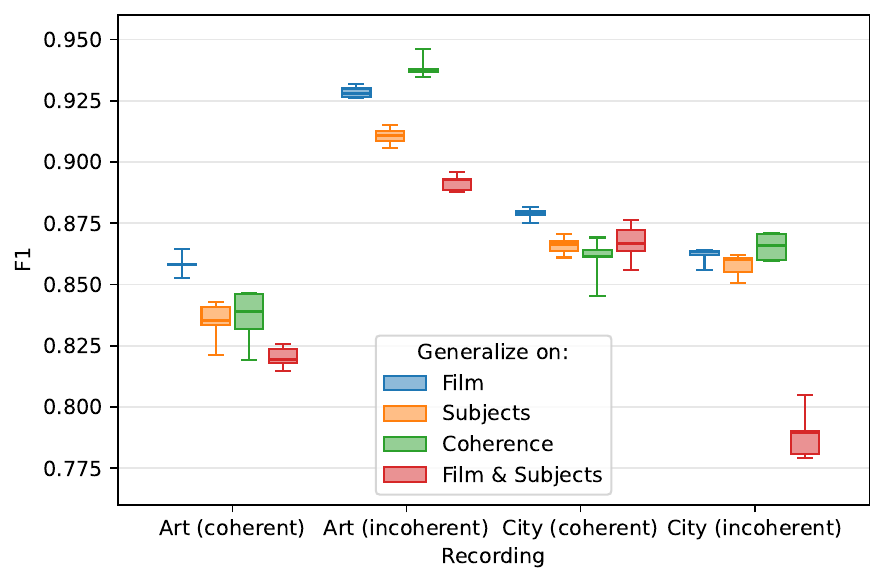}
        \caption{Box plot of the neural model's performance on the different EEG recordings, by the generalization objective.}
        \label{fig:trial_summary_plot_box}
    \end{figure*}

    We performed five independent training sessions for each setup.
    For each training session, we summarized performance by averaging the F1 scores for detected peaks across epochs 10 to 20.
    Figure~\ref{fig:trial_summary_plot_box} shows the box plot of the results.
    As expected, the most challenging training setup was the one in which the model had to generalize across both film and subject group. 
    This was likely due to both the amount of training data being halved and the harder generalization task.
    However, this setup prevented the detector from exploiting film- or subject-specific idiosyncrasies and tested whether the learned EEG signature of cuts transferred across stimulus material.
    The remaining experimental results in this section are based on this training setup.
    The difference in performance between the two films was also notable.
    Although all training setups achieved higher F1 scores when tested on the incoherent version of \emph{Art}, this pattern did not hold for \emph{City}.
    
    \paragraph{Error Analysis}
        The primary goal of EEG-based cut detection was to test whether TRPs triggered by visual and cognitive events are robust enough relative to noise to support cut identification.
        At the same time, detection errors, especially false positives, can reveal the types of patterns and narrative events that share EEG characteristics with cinematic cuts.
        Therefore, these errors may help to clarify what kinds of stimuli TRPs capture from the EEG.
        Upon initial inspection, false negatives did not appear to form a random subset: at least some corresponded to cuts in which the audiovisual transitions were too subtle,
        or in which transients occurred in rapid succession, making the signal difficult to resolve.
        For example, cuts corresponding to camera zoom or angle shifts seemed hard to notice.
        Likewise, false positives did not seem to appear at random either: many were auditory and/or visual transitions in the film, such as a light switching on or off, a picture being shown on a computer monitor,
        or a change of framing where a narratively important object or person appears.
        These may share common sensory or narrative features with true cuts.

        \begin{table}[h!]
            \centering
            \resizebox{\textwidth}{!}{
            \begin{tabular}{l|cc|cc|c}
                \noalign{\hrule height 1pt}
                \textbf{Film} & \multicolumn{2}{c|}{\textbf{\emph{Art}}} & \multicolumn{2}{c|}{\textbf{\emph{City}}} &  \\
                \textbf{Narrative} & \textbf{Coherent} & \textbf{Incoherent} & \textbf{Coherent} & \textbf{Incoherent} & \textbf{\%} \\
                \noalign{\hrule height 1pt}
                True positive           & 123 & 158 & 147 & 118  &  \\
                \hline
                False positive          & 4   & 28  & 17   & 5   &  \\
                \quad Transient         & 0   & 12  & 8    & 5   & 46.3\% \\
                \quad Significant event & 4   & 13  & 5    & 0   & 40.7\% \\
                \quad Undetermined      & 0   & 3   & 4    & 0   & 13.0\% \\
                \hline
                False negative          & 44  & 7   & 28   & 59  &  \\
                \quad Weak transient    & 7   & 0   & 8    & 10  & 18.1\% \\
                \quad Sub-window        & 12  & 5   & 16   & 28  & 44.2\% \\
                \quad Undetermined      & 25  & 2   & 4    & 21  & 37.7\% \\
                \noalign{\hrule height 1pt}
            \end{tabular}
            }
            \caption{Cut detection outcomes by EEG recording, manual error assessment.}
            \label{tab:detection_outcomes_gen_on_film_and_subj}
        \end{table}

        In order to analyze the errors by the above aspects, we categorized them manually (Table~\ref{tab:detection_outcomes_gen_on_film_and_subj}).
        This error analysis used all annotated cinematic cuts, rather than only the matched subset used for the TRP comparisons.
        False positives were classified as \emph{transients} when an intense auditory or visual transient was present other than a cinematic cut,
        as \emph{significant events} when the detection also coincided with a narratively relevant event, or as \emph{undetermined} otherwise.
        False negatives were classified as \emph{weak transients} when the audiovisual transient associated with the cut was subtle (e.g. cut involved only a change of camera angle),
        as \emph{sub-window} when a neighboring cut was present within a $\qty{1}{\second}$ window,
        as \emph{undetermined} when neither of these applied.
        Overall, the error structure suggests that the detector learned a physiologically meaningful cut-related response rather than an arbitrary mapping from EEG fluctuations to annotations.
        The majority of false negatives (62.3\%) occurred when the cinematic transition was perceptually weak or temporally crowded by neighboring cuts,
        whereas false positives mostly coincided (87\%) with strong audiovisual changes or narratively meaningful events.
        This pattern supports the interpretation that the detector was sensitive to EEG activity resembling event-boundary processing,
        even when such processing was not always tied to a formally annotated cinematic cut.
    
\begin{figure*}[h!]
             \centering
             \begin{subfigure}[b]{.49\textwidth}
                \centering
                \includegraphics[width=\textwidth]{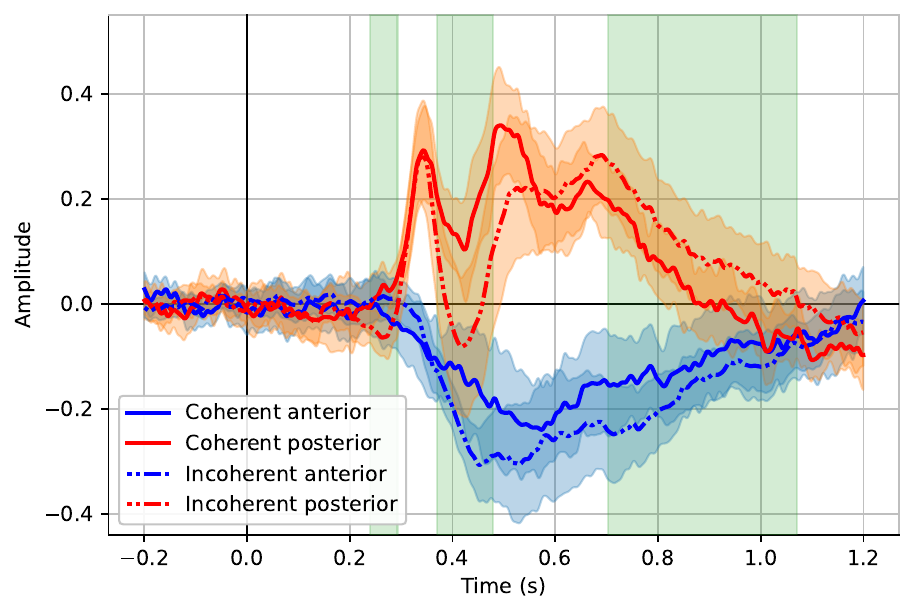}
                \subcaption{\emph{Art}}
                \label{fig:Clutsering_Art_l_v_nl_peak_detection}
             \end{subfigure}
             \hfill
             \begin{subfigure}[b]{.49\textwidth}
                \centering
                \includegraphics[width=\textwidth]{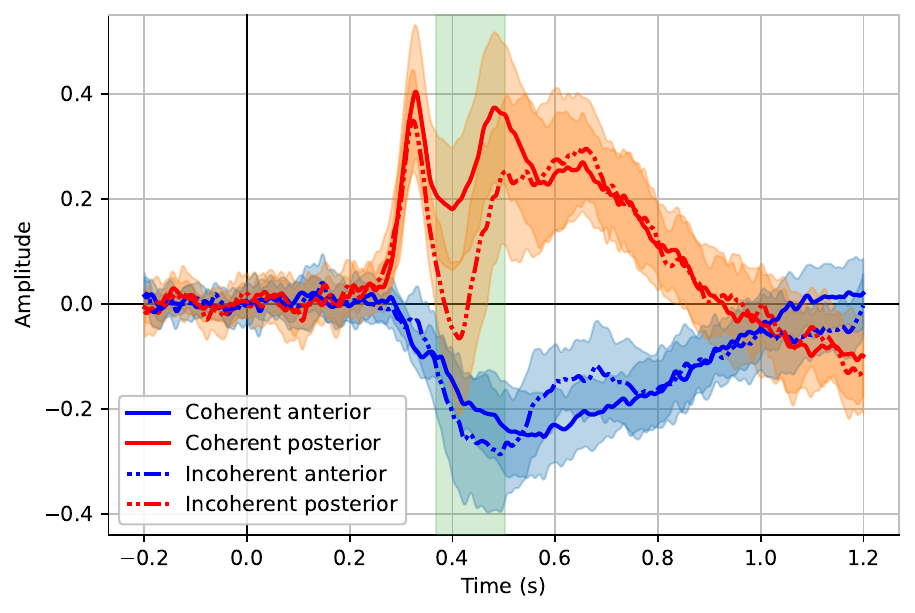}
                \subcaption{\emph{City}}
                \label{fig:Clutsering_City_l_v_nl_peak_detection}
             \end{subfigure}
             \caption{Anterior and posterior average TRPs of the detected cuts within the coherent and incoherent narrative versions of the same film.
             Green bars indicate significant clusters.}
            \label{fig:Clutsering_DL}
        \end{figure*}

    \paragraph{TRP Validation}
        To validate the efficacy of the model outputs, we repeated the TRP analysis from Section~\ref{sec:TRP} using the detected cuts (both true and false positives).
        The accuracy of the recovered TRPs is remarkable, considering that the model was blind to the cut-segmentation of the films.
        Figure~\ref{fig:Clutsering_DL} shows the resulting TRPs, and Table~\ref{tab:art_city_clusters_peak_detection} lists the cluster permutation test results.
        Most importantly, similar clusters were recovered as in the hand-annotated TRPs (Figure~\ref{fig:Clutsering_Gold} and Table~\ref{tab:art_city_clusters}) with marginally smaller confidence (larger $p$-values).
        These results not only corroborate the model's accuracy in finding cuts, but also highlight its temporal precision, which was essential for obtaining TRP waveforms with clear ERP-like features.
        Additionally, temporal PCA based on the detected cut TRPs yielded similar results to PCA based on the original cuts (Figure~\ref{fig:PCA_temporal_DL} in the Appendix).

        \begin{table}[h!]
            \centering
            \resizebox{\textwidth}{!}{
            \begin{tabular}{l|cccc}
            \noalign{\hrule height 1pt}
            Film & $t_{\min}$ & $t_{\max}$ & $p$-value & Channels \\
            \noalign{\hrule height 1pt}
            \emph{Art}  & 0.240 & 0.293 & 0.0128 & \footnotesize{\texttt{CP1,O1,O2,Oz,P3,P4,P7,P8,PO3,PO4,Pz}}\\
            \emph{Art}  & 0.369 & 0.479 & 0.0045 & \footnotesize{\texttt{AF3,AF4,C3,CP1,CP5,F3,F7,FC5,Fp1,Fp2,P3,P7,T7}}\\
            \emph{Art}  & 0.703 & 1.070 & 0.0025 & \footnotesize{\texttt{CP1,CP2,Cz,F3,F7,FC1,FC5,P3,P4,PO4,Pz}}\\
            \hline
            \emph{City} & 0.367 & 0.502 & 0.0053 & \footnotesize{\texttt{CP1,CP5,CP6,F8,FC6,O1,O2,Oz,P3,P4,P7,P8,PO3,PO4,T7,T8}}\\
            \noalign{\hrule height 1pt}
            \end{tabular}
            }
            \caption{Significant clusters for the detected cuts in \emph{Art} and \emph{City}.}
            \label{tab:art_city_clusters_peak_detection}
        \end{table}
    
    \paragraph{Classifying Narrative Coherence}
        Finally, we asked whether narrative coherence could be decoded from single-subject EEG responses.   
        To address this question, we computed subject-level TRPs separately for each participant and each film condition by averaging the cut-locked EEG segments of that participant.
        Each subject-level TRP was then cropped to the $\qtyrange{200}{1000}{\milli\second}$ post-cut interval, and the resulting $32 \times 409$ channel-by-time matrix was flattened into a single feature vector.
        We used these feature vectors to classify whether the corresponding TRP came from the coherent or incoherent version of the film.
        Classification was evaluated with leave-one-subject-out cross-validation, so that the subject being tested was never included in the training set.
        Because each subject contributed only one TRP per film, the classifier could not rely on repeated samples from the same subject within a film-specific dataset.
        Thus, successful classification would indicate that subject-level TRPs retained information about narrative coherence, rather than merely reflecting subject-specific idiosyncrasies.
        We repeated the analysis using both the original manually annotated cuts and the cut events detected by the DNN model.
        In the manually annotated version, cuts were matched one-to-one across coherent and incoherent conditions.
        This ensured that the classifiers could not rely on systematic differences in the immediate post-cut sensory input.
        
        \begin{table}[h!]
            \centering
            \begin{tabular}{l|cc|cc}
            \noalign{\hrule height 1pt}
            \textbf{Film} & \multicolumn{2}{c|}{\textbf{\emph{Art}}} & \multicolumn{2}{c}{\textbf{\emph{City}}} \\
            \textbf{Cut Event}  & \textbf{Original} & \textbf{Detected} & \textbf{Original} & \textbf{Detected} \\
            \noalign{\hrule height 1pt}
            Gradient Boost      & 0.71 & 0.74 & 0.82 & 0.74 \\
            LDA                 & 0.85 & 0.74 & 0.74 & 0.77 \\
            Logistic Regression & 0.82 & 0.85 & 0.85 & 0.77 \\
            Random Forest       & 0.88 & 0.85 & 0.74 & 0.71 \\
            Ridge Classifier    & 0.79 & 0.91 & 0.77 & 0.77 \\
            SVM                 & 0.82 & 0.85 & 0.82 & 0.77 \\
            XGBoost             & 0.82 & 0.74 & 0.77 & 0.79 \\
            \hline
            \textbf{Average}    & \textbf{0.815} & \textbf{0.811} & \textbf{0.786} & \textbf{0.756} \\
            \noalign{\hrule height 1pt}
            \end{tabular}
            \caption{Classification accuracy scores of narrative coherence based on true cuts and DNN-detected cut-like events.
            The numbers represent accuracy scores achieved by different ML models using leave-one-subject-out cross-validation.
            }
            \label{tab:classifier_scores}
        \end{table}

        Narrative coherence was classified using different machine learning (ML) approaches.
        As Table~\ref{tab:classifier_scores} indicates, average classifier performance decreased only marginally when detected cuts were used instead of original cut annotations.
        The classifiers correctly identified whether subject-level TRPs were derived from coherent or incoherent narratives for the majority of held-out subjects.
        Although the average accuracy of the ML models was $>75\%$, substantially above the empirical chance level, 
        the main point of this analysis was not to maximize classification performance.
        Instead, the aim was to demonstrate the efficacy of this method to extract subject-level cognitive information from EEG, and to further corroborate the quality of the DNN-based cut detections.

    \paragraph{Ablations}
        A baseline comparison is reported in \ref{sec:ablation}.
        In brief, the main DNN detector outperformed the anterior--posterior divergence, TRP-template-fit, and correlation-based baselines in the film-generalization setting.
        Importantly, the DNN localized cut onsets more accurately: on average, true positives were within one frame of the true cut onsets, while simpler peak-detection baselines approached the timing error expected from random detections within the acceptance window.
        Corroborating that temporal precision was critical for recovering high-quality TRPs from detected events.

\newpage
\section{Discussion} 
    EEG is the most accessible noninvasive tool for studying the brain's macroscopic electric potentials, which correlate with cognitive functions at millisecond precision.
To unlock EEG's full potential in diverse experimental settings \citep{wang2018concurrent, SONKUSARE2019699, lau2019mobile, kaushik2022decoding}
and access deeper layers of information processing in the brain we need to improve the sensitivity and specificity of the data processing method.
Even when extrinsic noise is managed, the biggest challenge is the large-amplitude intrinsic noise frequently obscuring the relationship between neuronal responses, particularly when stimuli are continuous and events occur at irregular intervals.
Conventional ERP analysis addresses this by averaging time-locked responses across repeated stimulus presentations \citep{luck2014introduction}, but this repetition renders most experiments unnatural and contrived.
One conventional solution developed to overcome this limitation was single-trial analysis, replacing waveform averaging with spectral or spatial filters and classifiers \citep{BLANKERTZ2011814}.
Instead, we developed an alternative approach to extract ERP-like transition-related potentials (TRPs) from continuous EEG recorded while subjects watched short films, time-locked to well-defined sensory events, such as cinematic cuts.

The first part of the study demonstrated that TRPs reproduce essential cognitive components of ERPs, specifically P2, P3 and N4 (Figure~\ref{fig:Gold_AP}). 
Among those, we showed that TRPs are systematically modulated by narrative context. 
Group-level TRPs drawn from EEGs recorded while participants watched scene-scrambled versions of films, exhibited reproducibly different characteristics relative to the TRPs computed from the EEGs while watching the original version.
Specifically, we observed significant modulation of P2 and P3 in the posterior and N400 amplitudes in the anterior electrodes consistent with the coherent-incoherent versions (Figure~\ref{fig:Clutsering_Gold}).
These effects were controlled by the counterbalanced allocation of participants across films and by one-to-one scene matching, which ensured identical audiovisual input in the post-cut TRP source-windows.
Furthermore, the TRP differences aligned with participants' subjective assessments, as scene-scrambled films were consistently rated as less narratively engaging.
The observed differences were further corroborated by temporal PCA, which separated two main components underlying the modulation of TRPs (Figure~\ref{fig:PCA_temporal}).
The first and strongest principal component, explaining $\approx90\%$ of signal variance, captured a broadly distributed posterior positivity: identified with the P2 and P3 components.
It also confirmed that the large topographic separation between anterior negativity and posterior positivity was the dominant characteristic of cut-elicited TRPs in general, and was not specific to the narrative differences between the films.
In contrast, the second principal component displayed a pronounced difference between coherent and incoherent narratives, with maximal expression over parietal and temporal areas.
Together, these findings support two conclusions: first, that cuts set a cascade of cognitive components into motion, originating over occipital cortex and followed by the evolution of a sustained frontal negativity; and second, that the modulation of these components by visual transients carries the fingerprints of highly abstract cognitive processing, specifically narrative coherence, a key determinant of film comprehension and, more broadly, of how meaning is constructed during continuous real-world experience.

It is worth to contextualize these findings within the ERP literature.
Regarding the positive components of the TRPs, both P2 and P3 are associated with the reduction of subjective uncertainty.
P2 \citep{luck1994electrophysiological} is commonly evoked in auditory oddball and visual priming paradigms, and is most pronounced centrally in auditory speech perception tasks \citep{signoret2013similarities},
with a more attenuated expression in visual paradigms \citep{capilla2016retinotopic}.
Its manifestation in our experiment was nonetheless evident: given that P2 is generally associated with novelty \citep{10.1093/cercor/bhm111},
it was expected to be most prominent following a cut when the new scene is less predictable.
Likewise, the P3 component is sensitive to the information-theoretic surprise value of a stimulus, with P3a over frontal and central areas reflecting attentional engagement \citep{Polich2003} and P3b,
with its posterior maximum, indexing stimulus unexpectedness \citep{Donchin-Q-surprise}.

Two competing hypotheses can be articulated regarding the P3b in our experiment.
The first predicts a more prominent P3b in incoherent-narrative EEGs, since the randomized sequence of shots produces a higher rate of unexpected transitions.
The second, however, draws on the relationship between P3b amplitude and information-theoretic unexpectedness,
as captured by Rényi entropy \citep{renyi1961entropy}, and leads to the opposite prediction.
In a coherent film, predictable transitions predominate, meaning that the rare unpredictable cut carries greater surprise and may therefore elicit a larger P3b than any individual transition in an incoherent film, where all cuts are equally unexpected and no strong prior exists to be violated.
This dependence on baseline stimulus probability is precisely what Rényi entropy formalizes, and trial-by-trial P3 modulation by such prior probabilities has been demonstrated in Bayesian paradigms \citep{mars2008trial}.

In addition to the positive components, we observed a prominent anterior negativity spanning from $250$ to $\qty{1000}{\milli\second}$, which we associate with the cognitive process of anticipation resolution.
This frontally-dominant negative potential is consistent with feedback-related negativity documented in the literature \citep{maruyama2026modulation}.
Although N400 is most commonly associated with linguistic tasks, it is not strictly a semantic component; it arises in any task involving the processing of meaning or conceptual information \citep{ganis1996search}.
This has been demonstrated in image-based sentence completion paradigms, where the final word is replaced by a predictable or unpredictable image rendering the sentence semantically congruous or anomalous, respectively \citep{nigam1992n400,ganis1996search}.
To our knowledge, the only prior study to elicit an N400-like response using video stimuli is that of \citet{sitnikova2003semantic}, who presented contextually congruous and incongruous film endings.
They found that incongruous endings evoked an enhanced anterior negativity resembling N400, as well as a more positive late positivity relative to congruous endings of short video presentations of human interactions with objects.
This pattern is perfectly concordant with our own finding that cuts demarcating incoherent narratives elicit larger N400 deflections and increased late P3b positivity.
Notably, the above referenced study called for exactly the paradigm we employed (p. 163):
``\emph{To get a better estimate of the timing of semantic integration in videos, it might be useful to collect ERPs to critical items that have a clear point of appearance in the scene (e.g., at a scene change).}''
Our experiment directly answers that call.

Next we turned the task around and approached it from the opposite direction.
Leveraging the power of sequence-processing artificial neural networks, we trained DNNs to exploit the characteristic EEG signature of TRPs and localize cinematic cuts in continuous EEG based solely on their electrographic signatures (Figure~\ref{fig:cut_detection}).
More specifically, we developed a DNN-based cut-detection methodology that located the vast majority of cinematic cuts from group-averaged EEG signals with a low error rate, and most importantly, with a sub-frame onset accuracy ($<\qty{40}{\milli\second}$).
With the exception of a single outlier, cut detection achieved consistently strong performance ($0.8 < \mathrm{F1} < 0.95$; Figure~\ref{fig:trial_summary_plot_box});
even in the most challenging data partitioning scenario, when generalizing across both, film and participant group, using only one group-averaged EEG recording for training.
Nevertheless, performance differences across all generalization conditions remained within $\Delta \mathrm{F1} = 0.05$, underscoring the robustness of the learned representations.
Although this approach has practical value as a method for simply detecting cinematic cuts only to a limited extent,
failures of the detector may point to events with cognitive properties similar to those of cuts.
    
Thus, investigating the detection errors proved highly informative. 
The majority of false negatives (62.3\%) were accounted for by cuts involving insufficiently salient visual transients (such as transparent transition), or by cuts occurring in rapid succession within the 1-second window during which multiple detections were suppressed.
False positives, by contrast, were of considerable interest, as they shed light on the cognitive underpinnings of cut detection.
Conceptually, a false positive arises when an EEG pattern shares the electrographic fingerprint of a TRP-template without being preceded by an actual cut.
Analyzing the false positives revealed a general pattern.
Some false positives coincided with audiovisual transients that were very similar to cuts while technically not being cinematic cuts, such as a light in the room being switched on/off, or a camera shutter effect.
Naturally, the transients associated with false positives (87\%), were on a spectrum in terms of intensity.
A general trend we found was that some of the weaker transients were also associated with narratively important moments, and this association was a likely factor in their detection.
For example, in one scene, on his computer monitor, the protagonist reviews photos he took earlier, and the transitions between these photos resemble cinematic cuts, but impact only a portion of the frame.
Out of the four pictures, the one that coincided with a false positive detection was the most narratively important, a portrait of the girl the protagonist is interested in.
While such events may lack a strong audiovisual transient and thus the early sensory components might have been too weak, their late cognitive components -- P2, P3, and N4 -- were nonetheless strong enough to be triggered by narrative events similar to cuts with respect to demanding an update to the viewer's internal model of the story.
The coincidence of false positives with such narrative events suggests that the DNN had learned to generalize TRP features beyond literal cuts, effectively detecting cognitively salient moments in the continuous EEG.

The timing precision of the detector was further demonstrated by the fact that detected events could be used directly to generate TRP waveforms that retained the main narrative-context-dependent differences observed for the original cut annotations (Figure~\ref{fig:Clutsering_DL}).
In addition, a participant-level classification of narrative coherence (inferring from individual TRPs if a participant watched the coherent or the incoherent version of the film), 
suffered only a marginal decrease in accuracy when TRPs were generated from the algorithmically detected cuts rather than from the original annotations.
These results appear even more compelling in light of the errors produced by the detector: many false negatives occurred in challenging cases, such as fast-paced cuts or weak audiovisual changes,
whereas many false positives coincided with strong audiovisual transients, often with narrative significance.
Taken together, these results provide strong evidence that the DNN learned to identify narratively significant cognitive events from EEG waveform signatures derived from cinematic cuts.

\section{Conclusion}
    Our findings collectively demonstrate that narrative comprehension leaves a measurable electrographic signature at moments of event transition, manifest as transition-related potentials (TRPs).
    Moreover, by leveraging the structure of TRPs, a DNN can be trained to extract analogous signatures from continuous EEG group-averages by capturing events that share physical or cognitive attributes with cinematic cuts or other well-defined stimulus boundaries.
    Furthermore, with the appropriate algorithmic approach, these detections are sufficiently accurate to automatically reproduce high-quality TRP waveforms.
    Importantly, these results indicate, that a robust mapping exists between continuous EEG dynamics and the timing of externally defined narrative transitions.
    Such electrographic signatures offer a principled means of parsing continuous EEG for cognitive events representing episodes of information updating, uncertainty reduction, or the resolution of suspense:
    hallmarks of the dramatic architecture of produced films and, more broadly, of natural experience \citep{DiniENEURO.0484-22.2023}.
    This discovery opens the door to detecting meaningful transitions in other continuous stimuli, where event boundaries may not be directly observable or cannot be annotated by hand with comparable precision.
    Such applications would extend the TRP framework beyond cinematic cuts and enable the analysis of information processing across a broader range of continuous, naturalistic experiences.
    Future efforts will be devoted to test the generalization of TRPs and to parse real-time individual EEG.

\section{Limitations}
    Several limitations should be considered when interpreting these results.
First, the experiment used only two short films with clear narrative structure and relatively sparse dialogue.
The generality of the findings should therefore be tested on a wider range of films, genres, editing styles, and audiovisual materials.
Second, the DNN detector operated on subject-averaged EEG.
It should therefore be interpreted as a population-level detector rather than a real-time single-subject detector.
Future work should test whether subject-adaptive models can recover comparable event timing from individual EEG recordings.
Third, the interpretation of false positives and false negatives relied partly on manual categorization.
Combining EEG-based detection with richer stimulus annotations, including auditory changes, object appearances, semantic shifts, and narrative turning points were out of scope of this study, and would be subject of further investigation.
Fourth, due to the small number of experimental subjects, the TRP results are based on signals averaged across both individual cuts and participants.
Future studies aiming to analyze responses to individual cuts should therefore use substantially larger participant samples.
Finally, the analyses reported here are scalp-level analyses.
They identify reproducible temporal and topographic patterns in the EEG, but they do not localize the neural generators of these effects.
Source-level analyses or multimodal recordings would be needed to determine the neural systems underlying the observed TRP components.

\section*{Ethics Statement}
    All participants provided written informed consent after receiving a full explanation of the experimental procedure.
    The study was approved by the United Review Committee for Research in Psychology (Hungary, EPKEB 2021-123).
    The study was conducted in accordance with the Declaration of Helsinki.
    Participants received no financial compensation.

\section*{Data and Code Availability Statement}
    The EEG dataset is available at Mendeley Data \citep{csanady2026continuous}.
    Code is available at GitHub \citep{csanady2026eegcuts}.
    
\section*{Declaration of Competing Interest}    
    The authors declare that they have no conflict of interest. 

\section*{Credit Authorship Contribution Statement}
    Bálint Csanády: \emph{conceptualization, methodology, software, formal analysis, investigation, writing -- original draft, data visualization}.
    Péter Vedres: \emph{conceptualization, methodology, formal analysis, investigation}.
    Kristóf Zsolt Makó: \emph{software, investigation}.
    Orsolya Papp-Zipernovszky: \emph{research design, psychometric assessment}.
    Márta Volosin: \emph{research design, investigation, EEG data recording}.
    Dávid Apagyi: \emph{software, investigation}.
    András Lukács: \emph{supervision, methodology}.
    András Bálint Kovács: \emph{conceptualization, research design, methodology}.
    Zoltán Nádasdy: \emph{conceptualization, methodology, supervision, writing -- review and editing}.
   
\section*{Acknowledgments}
    This research was supported by the EKÖP-25-4-I-ELTE-1093 University Research Fellowship Program of the Ministry for Culture and Innovation from the Source of the National Research, Development and Innovation Fund.
    The research was additionally supported by the Hungarian National Research, Development and Innovation Office within the framework of the Thematic Excellence Program 2021 -- National Research Sub programme: ``Artificial intelligence, large networks, data security: mathematical foundation and applications'' and the Artificial Intelligence National Laboratory Program (MILAB).
    This research was also funded by the OTKA research funding program under project number NKFI-138108.

\bibliographystyle{elsarticle-harv}
\bibliography{bibliography}

\newpage \appendix
\section{Ablation Studies}\label{sec:ablation}
    To gain insight into which aspects of the cut-related EEG waveforms the DNN cut detector is sensitive to,
we compared it with alternative basic cut-detection approaches informed by the observed TRP waveforms and the previous literature.
These detectors were designed to capture different properties of the EEG response expected around cuts.
They served as interpretable reference points for the DL approach and helped illustrate why accurate, temporally precise event detection is important for TRP analysis.

\begin{figure*}[h!]
    \centering
    \begin{subfigure}[b]{.49\textwidth}
        \centering
        \includegraphics[width=\textwidth]{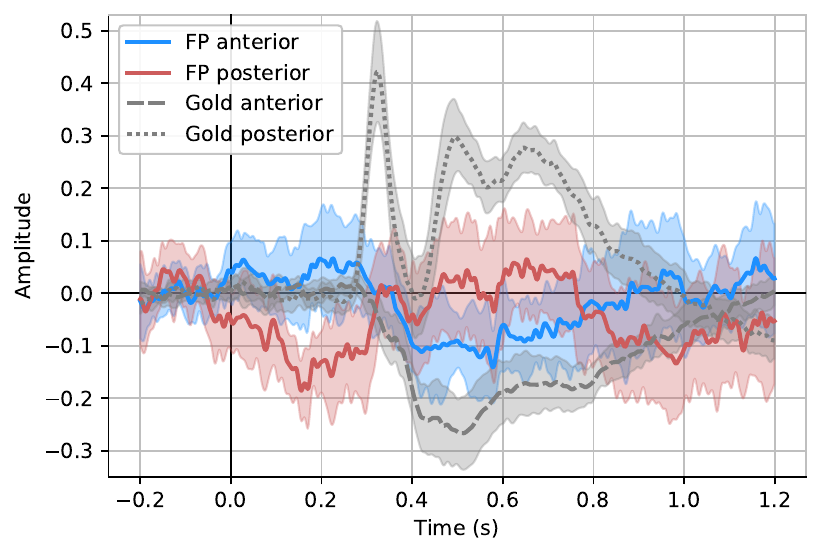}
        \subcaption{False Positives}
        \label{fig:gen_on_film_FP}
    \end{subfigure}
    \hfill
    \begin{subfigure}[b]{.49\textwidth}
        \centering
        \includegraphics[width=\textwidth]{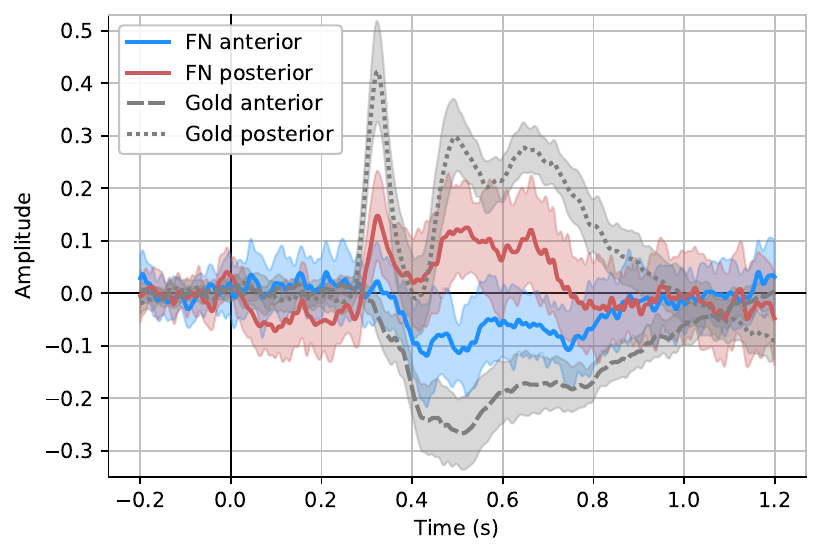}
        \subcaption{False Negatives}
        \label{fig:gen_on_film_FN}
    \end{subfigure}
    \caption{Anterior and posterior TRPs drawn from false positive and false negative triggers produced by the DNN model in the film-generalization setting. Signals averaged over all films.}
    \label{fig:gen_on_film_FPFN}
\end{figure*}

The TRP waveforms derived from the DNN model’s false-positive and false-negative detections revealed the following patterns (Figure~\ref{fig:gen_on_film_FPFN}).
First, anterior--posterior separation between $400$ and $800\,\mathrm{ms}$ appeared significantly less prominent in both the false positive and false negative TRPs, than in the TRPs generated from the original (gold) cuts. 
Second, the false positive TRP lacks the distinct components such as P2 in the posterior average, while the false negative TRP still resembles the expected TRP waveform.
Regarding the anterior--posterior divergence, when we shift the mean-correction interval from $\qtyrange{0}{200}{\milli\second}$ to $\qtyrange{100}{300}{\milli\second}$,
the false positive TRP shows stronger anterior--posterior separation.

These differences between the false positive and false negative TRPs may arise because false positives are model-detected events 
and may therefore be less precisely time-locked, 
whereas false negatives are a subset of the true cut annotations.
Nevertheless, these patterns motivate two basic cut-detection strategies. 
First, if the DNN model is less likely
to detect cuts when the anterior--posterior separation is less prominent, whereas its false positives still display such a separation,
then a detector based on anterior--posterior divergence may provide a useful baseline for identifying film events that trigger cognitive updating.
In other words, by such a detector we asked to what extent the DNN model's performance could be explained by simply detecting anterior--posterior separation.
Second, if the model misses cuts that still express ERP-like features, while producing false positives that do not, then a detector based on faithful TRP-template matching is worth investigating as a potential alternative.

We defined the anterior--posterior (A--P) divergence detector as follows.
Electrodes were grouped into anterior\footnote{\texttt{Fp1, Fp2, AF3, AF4, F7, F3, Fz, F4, F8, FC5, FC1, FC2, FC6, C3, C4.}}
and posterior\footnote{\texttt{T7, T8, CP5, CP1, CP2, CP6, P7, P3, Pz, P4, P8, PO3, PO4, O1, Oz, O2.}} regions, and the signal within each region was averaged across channels.
The metric was computed within normalized (\qty{1}{\second}) windows as the signed root mean square difference between the anterior and posterior regional averages.
This score takes high values during periods in which the two regions diverge with the polarity pattern expected for cut-related responses.
We then applied peak detection to the resulting score signal.
Because the peaks corresponding to the strongest anterior--posterior separation occurred at a consistent latency after cut onset,
we corrected the detected peak times by shifting them backward by a fixed temporal offset.

The second basic approach, the TRP-fit detector was calculated by fitting the cut-locked TRP template on the continuous EEG signal.
We used cross-validation generalizing on films: the TRP template calculated from \emph{Art} was fitted to the EEG signal of \emph{City}, and vice versa.
As in the A--P divergence cut-detector, the signals were first averaged within the anterior and posterior electrode groups.
Separate TRP templates were then fitted to the anterior and posterior regional averages.
The resulting fit score was evaluated over time on the test signal.
Cut predictions were then obtained from this metric by peak-detection, as in the A--P divergence detector.
To make the comparison less sensitive to slow drifts, the matching was performed on the discrete difference of the signals rather than on the raw waveform.

\subsection{Correlation-Based Cut-Detectors}
    The correlation-based cut detectors were motivated by the assumption that salient events such as cinematic cuts,
    which may represent event boundaries in the film's narrative \citep{zacks2010cuttingroom,magliano2011continuityediting},
    evoke synchronized responses across viewers \citep{hasson2004intersubject,dmochowski2012correlated,ki2016attention,nastase2019measuring}.
    Under this assumption, cut-related time periods should be accompanied by increased similarity between the EEG signals of different subjects.
    The detectors therefore measure inter-subject similarity over time and use this signal to identify candidate cut locations.
    We compared multiple variants of the correlation-based cut detection.
    
    \begin{itemize}
        \item First, we measured the average inter-subject Pearson correlation within $\qty{0.5}{\second}$ rolling windows, applied peak detection to the resulting time series,
        and shifted the peaks to the left by a constant amount, similar to  the A--P divergence cut-detector.
    \end{itemize}
    Additionally, we trained the same peak-detection DNN architecture as discussed in Section~\ref{sec:cut_detect}, but instead of training on averaged EEG signals, we trained on correlation-derived signals:
    \begin{itemize}
        \item We trained one version on a 32-channel signal in which each channel contained the inter-subject Pearson correlation of the corresponding EEG channel within a $\qty{1}{\second}$ moving window.
        \item Moreover, we trained a detector on a two-channel signal in which the two channels corresponded to the average correlations within the anterior and posterior electrode groups.
    \end{itemize}

\subsection{Ablation Results}
    \begin{table}[p]
        \centering
        \resizebox{0.82\textwidth}{!}{
        \begin{tabular}{l|ccccc}
        \noalign{\hrule height 1pt}
        \textbf{Model} & \textbf{F1}  & \textbf{TP} & \textbf{FP} & \textbf{FN} & \textbf{Avg. distance} \\
        \noalign{\hrule height 1pt}
        DNN                & 0.863 & 139 & 16  & 28 & \qty{40}{\milli\second}  \\
        Correlation DNN    & 0.610 & 111 & 86  & 56 & \qty{90}{\milli\second}  \\
        Correlation DNN AP & 0.465 & 91  & 133 & 76 & \qty{132}{\milli\second} \\
        Correlation        & 0.472 & 94  & 137 & 73 & \qty{151}{\milli\second} \\
        A-P divergence     & 0.442 & 74  & 94  & 93 & \qty{147}{\milli\second} \\
        TRP-fit            & 0.413 & 99  & 214 & 68 & \qty{153}{\milli\second} \\
        No Skill           & \emph{0.278} & \emph{167} & \emph{868} & \emph{0} & \emph{156}\,\qty{}{\milli\second} \\
        \noalign{\hrule height 1pt}
        \end{tabular}
        }
        \caption{Evaluation metrics on \emph{Art coherent} by model.}
        \label{tab:model_metrics_Art_l}
    \end{table}
    \begin{table}[p]
        \centering
        \resizebox{0.82\textwidth}{!}{
        \begin{tabular}{l|ccccc}
        \noalign{\hrule height 1pt}
        \textbf{Model} & \textbf{F1} & \textbf{TP} & \textbf{FP} & \textbf{FN} & \textbf{Avg. distance} \\
        \noalign{\hrule height 1pt}
        DNN                & 0.891 & 151 & 14  & 23 & \qty{23}{\milli\second} \\
        Correlation DNN    & 0.707 & 116 & 37  & 59 & \qty{74}{\milli\second} \\
        Correlation DNN AP & 0.555 & 93  & 67  & 82 & \qty{107}{\milli\second} \\
        Correlation        & 0.559 & 109 & 106 & 66 & \qty{139}{\milli\second} \\
        A-P divergence     & 0.519 & 87  & 73  & 88 & \qty{155}{\milli\second} \\
        TRP-fit            & 0.481 & 117 & 195 & 58 & \qty{150}{\milli\second} \\
        No Skill           & \emph{0.355} & \emph{175} & \emph{636} & \emph{0} & \emph{156}\,\qty{}{\milli\second}  \\
        \noalign{\hrule height 1pt}
        \end{tabular}
        }
        \caption{Evaluation metrics on \emph{City coherent} by model.}
        \label{tab:model_metrics_City_l}
    \end{table}
    \begin{table}[p]
        \centering
        \resizebox{0.82\textwidth}{!}{
        \begin{tabular}{l|ccccc}
        \noalign{\hrule height 1pt}
        \textbf{Model} & \textbf{F1} & \textbf{TP} & \textbf{FP} & \textbf{FN} & \textbf{Avg. distance} \\
        \noalign{\hrule height 1pt}
        DNN                & 0.953 & 161 & 12  & 4  & \qty{29}{\milli\second}\\
        Correlation DNN    & 0.712 & 142 & 92  & 23 & \qty{60}{\milli\second}\\
        Correlation DNN AP & 0.509 & 110 & 157 & 55 & \qty{122}{\milli\second}\\
        Correlation        & 0.490 & 106 & 162 & 59 & \qty{144}{\milli\second}\\
        A-P divergence     & 0.516 & 89  & 91  & 76 & \qty{164}{\milli\second}\\
        TRP-fit            & 0.397 & 104 & 255 & 61 & \qty{163}{\milli\second}\\
        No Skill           & \emph{0.277} & \emph{165} & \emph{860} & \emph{0} & \emph{156}\,\qty{}{\milli\second}  \\
        \noalign{\hrule height 1pt}
        \end{tabular}
        }
        \caption{Evaluation metrics on \emph{Art incoherent} by model.}
        \label{tab:model_metrics_Art_nl}
    \end{table}
    \begin{table}[p]
        \centering
        \resizebox{0.82\textwidth}{!}{
        \begin{tabular}{l|ccccc}
        \noalign{\hrule height 1pt}
        \textbf{Model} & \textbf{F1} & \textbf{TP} & \textbf{FP} & \textbf{FN} & \textbf{Avg. distance} \\
        \noalign{\hrule height 1pt}
        DNN                & 0.868 & 145 & 11  & 33  & \qty{22}{\milli\second}\\
        Correlation DNN    & 0.710 & 114 & 29  & 64  & \qty{66}{\milli\second}\\
        Correlation DNN AP & 0.550 & 90  & 59  & 88  & \qty{110}{\milli\second}\\
        Correlation        & 0.529 & 106 & 117 & 72  & \qty{133}{\milli\second}\\
        A-P divergence     & 0.460 & 74  & 70  & 104 & \qty{154}{\milli\second}\\
        TRP-fit            & 0.508 & 113 & 154 & 65  & \qty{147}{\milli\second}\\
        No Skill           & \emph{0.381} & \emph{178} & \emph{575} & \emph{0} & \emph{156}\,\qty{}{\milli\second} \\
        \noalign{\hrule height 1pt}
        \end{tabular}
        }
        \caption{Evaluation metrics on \emph{City incoherent} by model.}
        \label{tab:model_metrics_City_nl}
    \end{table}

    We compared the basic cut detectors to the main DNN cut detector.
    All detectors were evaluated in the film-generalization cross validation scenario, meaning that where applicable,
    training was performed on both versions of one film and evaluation on both versions of the other film.
    As a reference, we also report a no-skill baseline, which provides an upper bound for a random detector constrained by the same minimum inter-peak distance.
    Because the minimum inter-peak distance was fixed to $\qty{0.625}{\second}$, and the acceptance window also spanned $\qty{0.625}{\second}$ in total,
    this baseline distributed detections uniformly every $\qty{0.625}{\second}$ throughout the recording.
    This construction achieves perfect recall and yields maximal F1 score among no-skill detectors whose successive detections must be at least $\qty{0.625}{\second}$ apart.
    Its purpose was to define a simple reference level against which the data-driven methods could be compared.
    
    \Cref{tab:model_metrics_Art_l,tab:model_metrics_City_l,tab:model_metrics_Art_nl,tab:model_metrics_City_nl} summarize the model performance across films and narrative versions.
    Among the models compared, the main DNN model detected cuts most accurately, achieving the highest F1 scores and the fewest errors across all conditions.
    The average distance (temporal error) of the detected true positive cuts by the DNN model was within one frame ($\qty{40}{\milli\second}$ at $25$ frames per second).
    The simple peak-detection-based models performed better than random (no-skill baseline),
    however, their average temporal errors were close to $\qty{156.25}{\milli\second}$, the value expected if detections were distributed within the acceptance windows uniformly at random.

    \begin{figure*}[h!]
        \centering
        \begin{subfigure}[b]{.49\textwidth}
            \centering
            \includegraphics[width=\textwidth]{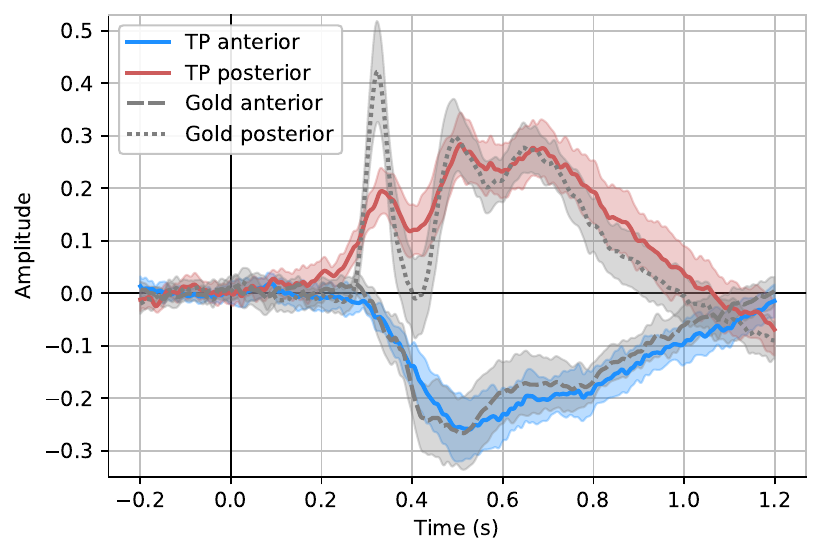}
            \subcaption{All 32 channel correlations.}
            \label{fig:gen_on_film_corr_32_TP}
        \end{subfigure}
        \hfill
        \begin{subfigure}[b]{.49\textwidth}
            \centering
            \includegraphics[width=\textwidth]{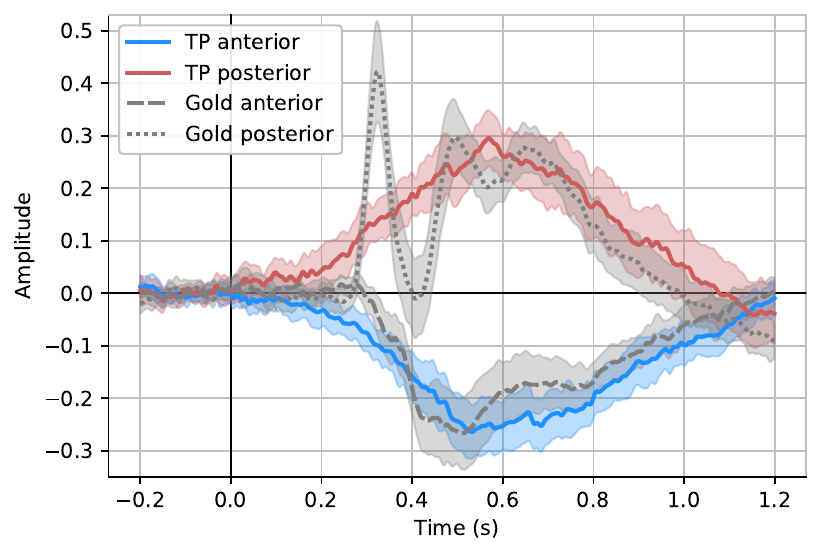}
            \subcaption{A--P group avg. correlations.}
            \label{fig:gen_on_film_corr_AP_TP}
        \end{subfigure}
        \caption{Anterior and posterior TRPs drawn from the true positive triggers by the DNN models trained on channel correlations. Models generalized on the film, signals averaged over all films.}
        \label{fig:gen_on_film_corr_TP}
    \end{figure*}

    Frame-accurate cut detection is crucial for reliably reproducing TRP waveforms.
    Figure~\ref{fig:gen_on_film_corr_TP} shows the TRPs generated from true positive detections of the DNN models trained on correlation data.
    The TRP generated from true positive detections of the $32$-channel correlation DNN model is already heavily attenuated at an average temporal error of $\qty{71.67}{\milli\second}$, which is less than two frames (Figure~\ref{fig:gen_on_film_corr_32_TP}).
    Moreover, at an average temporal error of less than three frames ($\qty{117.95}{\milli\second}$), the anterior--posterior correlation-based DNN model already produces a true positive TRP waveform with no noticeable ERP-like features apart from basic anterior--posterior divergence (Figure~\ref{fig:gen_on_film_corr_AP_TP}).

\subsection{Error Analysis}
    \begin{table}[h!]
        \centering
        \resizebox{\textwidth}{!}{
        \begin{tabular}{l|cc|cc|c}
            \noalign{\hrule height 1pt}
            \textbf{Film} & \multicolumn{2}{c|}{\textbf{\emph{Art}}} & \multicolumn{2}{c|}{\textbf{\emph{City}}} & \\
            \textbf{Narrative} & \textbf{Coherent} & \textbf{Incoherent} & \textbf{Coherent} & \textbf{Incoherent} & \textbf{\%} \\
            \noalign{\hrule height 1pt}
            True positive           & 139 & 161 & 151 & 145  & \\
            \hline
            False positive          & 16  & 12  & 14  & 11   & \\
            \quad Transient         & 5   & 5   & 13  & 7    & 56.6\% \\
            \quad Significant event & 9   & 6   & 1   & 2    & 34.0\% \\
            \quad Undetermined      & 2   & 1   & 0   & 2    & 9.4\%  \\
            \hline
            False negative          & 28  & 4   & 23  & 33   & \\
            \quad Weak transient    & 2   & 0   & 7   & 8    & 19.3\% \\
            \quad Sub-window        & 9   & 3   & 14  & 20   & 52.3\% \\
            \quad Undetermined      & 17  & 1   & 2   & 5    & 28.4\% \\
            \noalign{\hrule height 1pt}
        \end{tabular}
        }
        \caption{DNN cut detection outcomes by EEG recording in the film-generalization setting.}
        \label{tab:detection_outcomes_gen_on_film}
    \end{table}

    \begin{table}[h!]
        \centering
        \resizebox{\textwidth}{!}{
        \begin{tabular}{l|cc|cc|c}
            \noalign{\hrule height 1pt}
            \textbf{Film} & \multicolumn{2}{c|}{\textbf{\emph{Art}}} & \multicolumn{2}{c|}{\textbf{\emph{City}}} &  \\
            \textbf{Narrative} & \textbf{Coherent} & \textbf{Incoherent} & \textbf{Coherent} & \textbf{Incoherent} & \textbf{\%} \\
            \noalign{\hrule height 1pt}
            True positive           & 111 & 142 & 116 & 114 & \\
            \hline
            False positive          & 85  & 92  & 37  & 29  & \\
            \quad Transient         & 19  & 32  & 14  & 9   & 30.5\% \\
            \quad Significant event & 51  & 39  & 8   & 10  & 44.4\% \\
            \quad Undetermined      & 15  & 21  & 15  & 10  & 25.1\% \\
            \hline
            False negative          & 55  & 23  & 59  & 64  & \\
            \quad Weak transient    & 4   & 0   & 9   & 13  & 12.9\% \\
            \quad Sub-window        & 15  & 10  & 27  & 28  & 39.8\% \\
            \quad Undetermined      & 36  & 13  & 23  & 23  & 47.3\% \\
            \noalign{\hrule height 1pt}
        \end{tabular}
        }
        \caption{Cut detection outcomes by EEG recording for the 32-channel correlation DNN model in the film-generalization setting.}
        \label{tab:detection_outcomes_gen_on_film_corr}
    \end{table}
    
\begin{figure*}[h!]
        \centering
            \begin{subfigure}[b]{.49\textwidth}
            \centering
            \includegraphics[width=\textwidth]{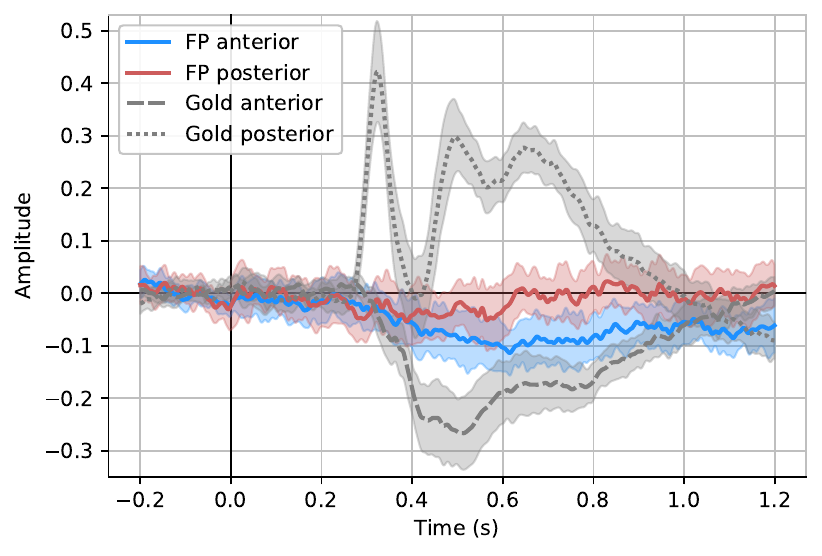}
            \subcaption{All 32 channel correlations.}
            \label{fig:gen_on_film_corr_32_FP}
        \end{subfigure}
        \hfill
        \begin{subfigure}[b]{.49\textwidth}
            \centering
            \includegraphics[width=\textwidth]{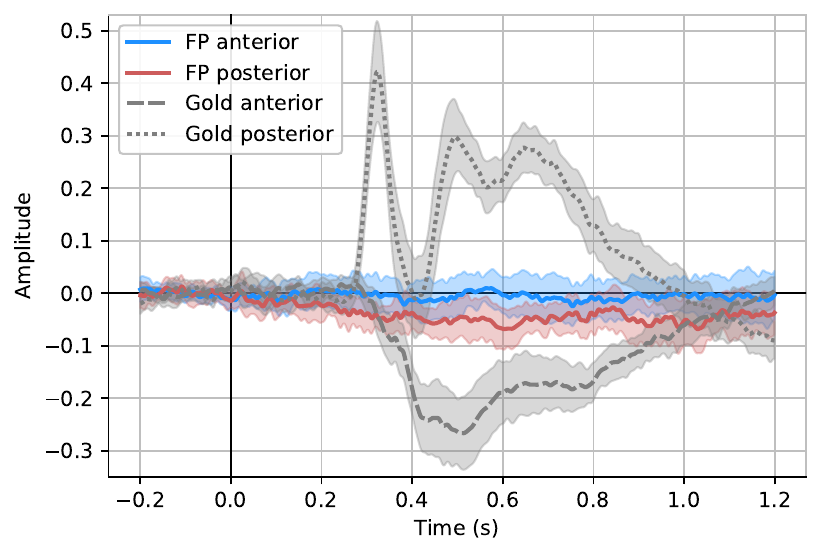}
            \subcaption{A--P group avg. correlations.}
            \label{fig:gen_on_film_corr_AP_FP}
        \end{subfigure}
        \caption{Anterior and posterior TRPs drawn from the false positive triggers by the DNN models trained on channel correlations. Models generalized on the film, signals averaged over all films.}
        \label{fig:gen_on_film_corr_FP}
    \end{figure*}

    Although the correlation-based cut detectors performed worse than the main DNN model and operated with lower temporal precision, we hypothesized that they might identify more narratively relevant false positives.
    \Cref{tab:detection_outcomes_gen_on_film,tab:detection_outcomes_gen_on_film_corr}
    compare the manual error-analysis results for the main DNN model and the 32-channel correlation-based DNN model.
    As anticipated, the correlation-based model identified a larger number of narratively relevant events.
    However, these events cannot be directly used in the standard TRP framework, because they do not exhibit strong ERP-like features when averaged around the detected peaks (Figure~\ref{fig:gen_on_film_corr_FP}).
    This suggests that correlation-based alternatives may be useful for discovering potentially interesting moments in the stimulus, but not directly for recovering meaningful event-related waveforms.
    Leveraging the additional potential of these significant false positive detections, would, therefore, require analysis methods beyond the scope of this study.

\section{Additional Figures}
    \begin{figure*}[h!]
    \centering
    \begin{subfigure}[b]{.49\textwidth}
        \centering
        \includegraphics[width=\textwidth]{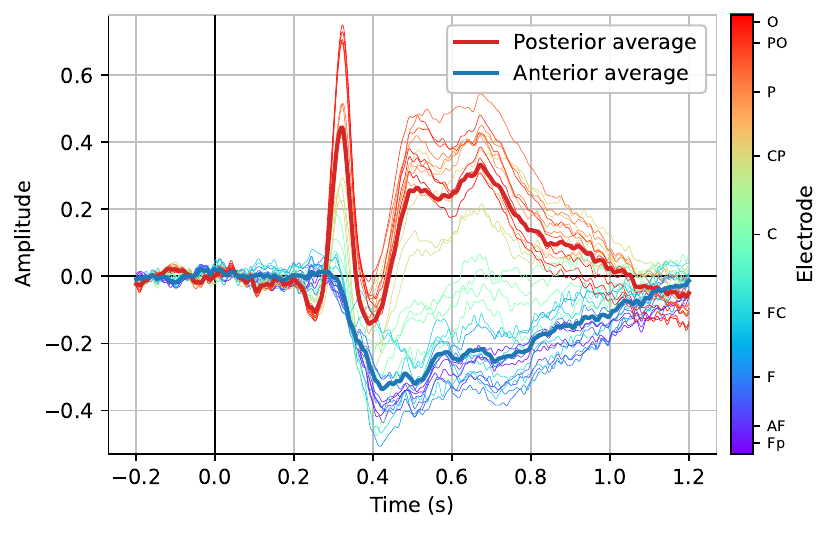}
        \subcaption{\emph{Art} (incoherent).}
        \label{fig:Art_Gold_AP}
    \end{subfigure}
    \hfill
    \begin{subfigure}[b]{.49\textwidth}
        \centering
        \includegraphics[width=\textwidth]{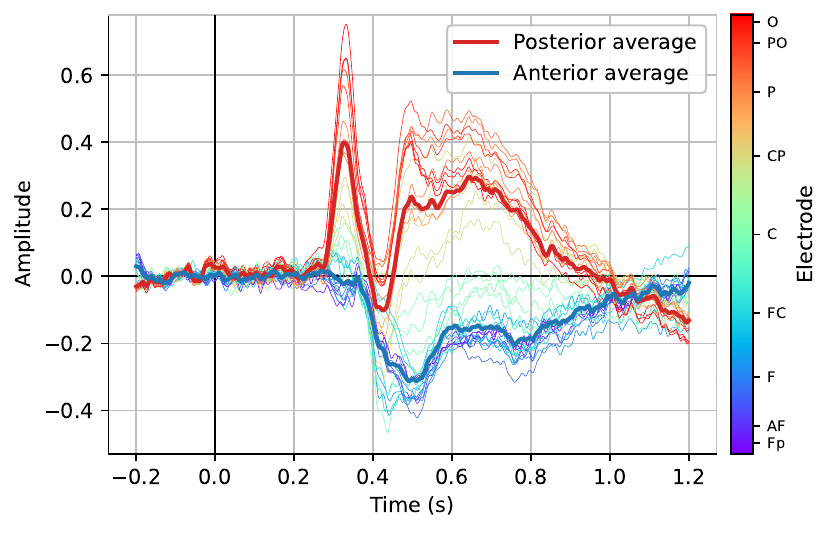}
        \subcaption{\emph{City} (incoherent).}
        \label{fig:City_Gold_AP}
    \end{subfigure}
    \caption{Butterfly plots of the average anterior and posterior TRP signals aligned to all cuts in the incoherent versions of the films.}
    \label{fig:Gold_AP_appx}
\end{figure*} 

\begin{figure*}[h!]
    \centering
    \begin{subfigure}[b]{.49\textwidth}
        \centering
        \includegraphics[width=\textwidth]{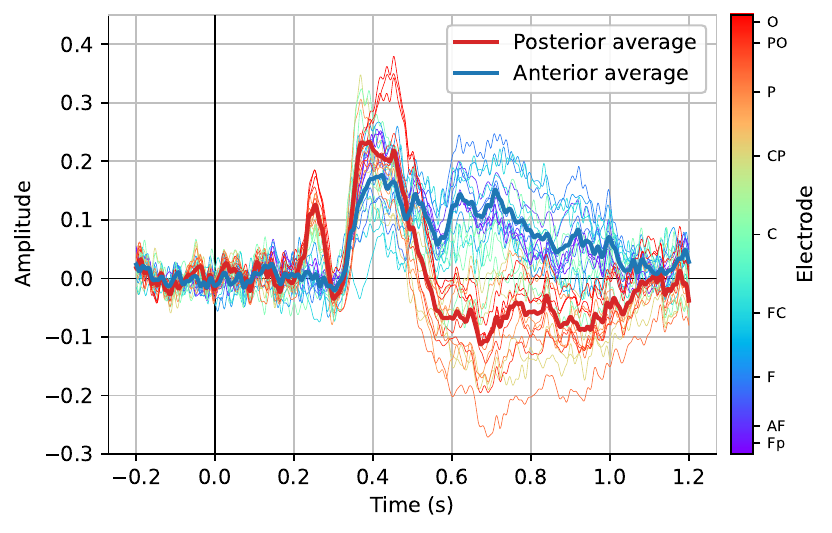}
        \subcaption{\emph{Art} ($\text{coherent}-\text{incoherent}$).}
        \label{fig:Art_Gold_AP_diff}
    \end{subfigure}
    \hfill
    \begin{subfigure}[b]{.49\textwidth}
        \centering
        \includegraphics[width=\textwidth]{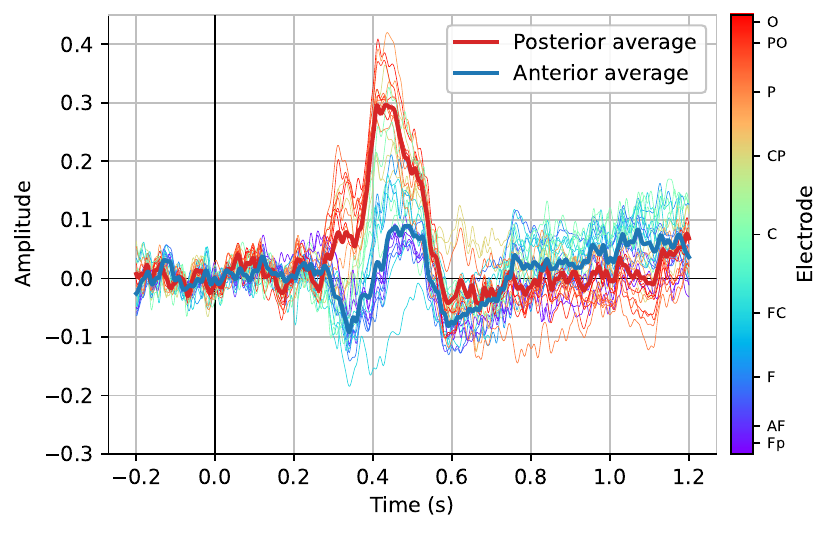}
        \subcaption{\emph{City} ($\text{coherent}-\text{incoherent}$).}
        \label{fig:City_Gold_AP_diff}
    \end{subfigure}
    \caption{Butterfly plots of the $\text{coherent}-\text{incoherent}$ average TRP signals.}
    \label{fig:Gold_AP_appx2}
\end{figure*} 

\begin{figure*}[h!]
     \centering
     \begin{subfigure}[b]{.242\textwidth}
        \centering
        \includegraphics[width=\textwidth]{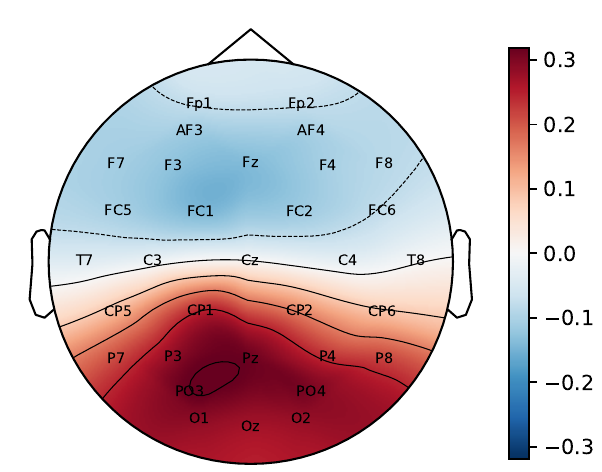}
        \subcaption{\emph{Art} eigenvector 1.}
     \end{subfigure}
     \hfill
     \begin{subfigure}[b]{.242\textwidth}
        \centering
        \includegraphics[width=\textwidth]{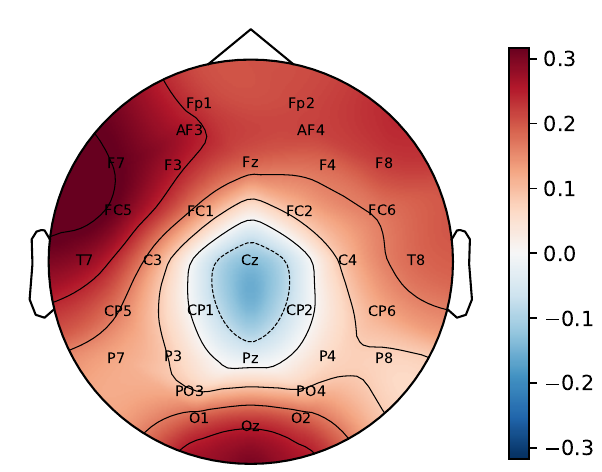}
        \subcaption{\emph{Art} eigenvector 2.}
     \end{subfigure}
     \hfill
     \begin{subfigure}[b]{.242\textwidth}
        \centering
        \includegraphics[width=\textwidth]{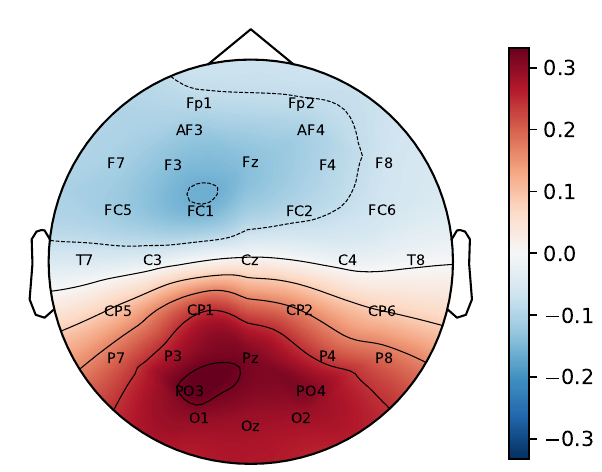}
        \subcaption{\emph{City} eigenvector 1.}
     \end{subfigure}
     \hfill
     \begin{subfigure}[b]{.242\textwidth}
        \centering
        \includegraphics[width=\textwidth]{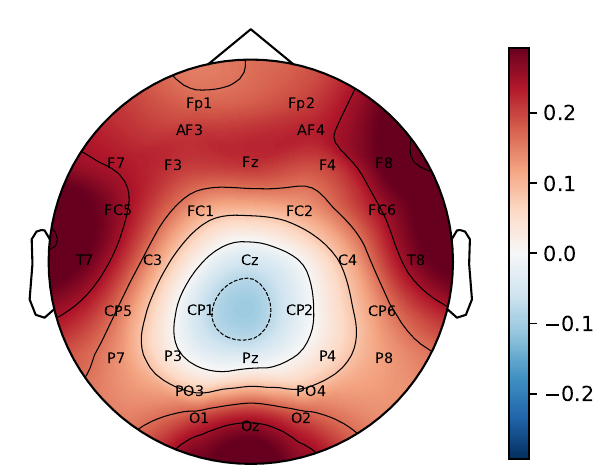}
        \subcaption{\emph{City} eigenvector 2.}
     \end{subfigure}
     \vskip 3mm
     \begin{subfigure}[b]{.49\textwidth}
        \centering
        \includegraphics[width=\textwidth]{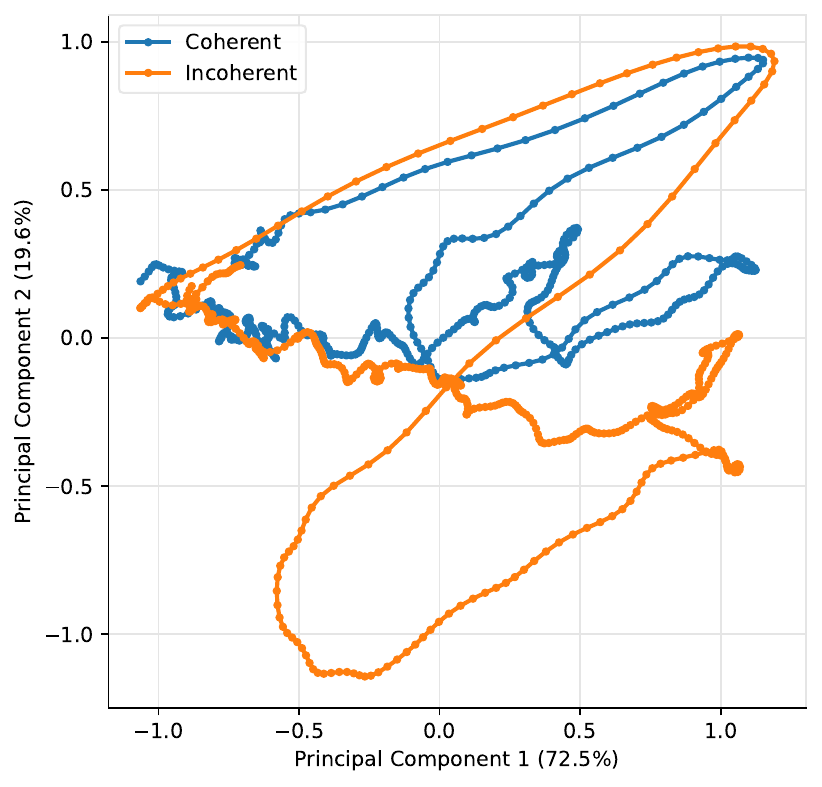}
        \subcaption{\emph{Art} principal components.}
        \label{fig:Art_PC12_parametric}
     \end{subfigure}
     \hfill
     \begin{subfigure}[b]{.49\textwidth}
        \centering
        \includegraphics[width=\textwidth]{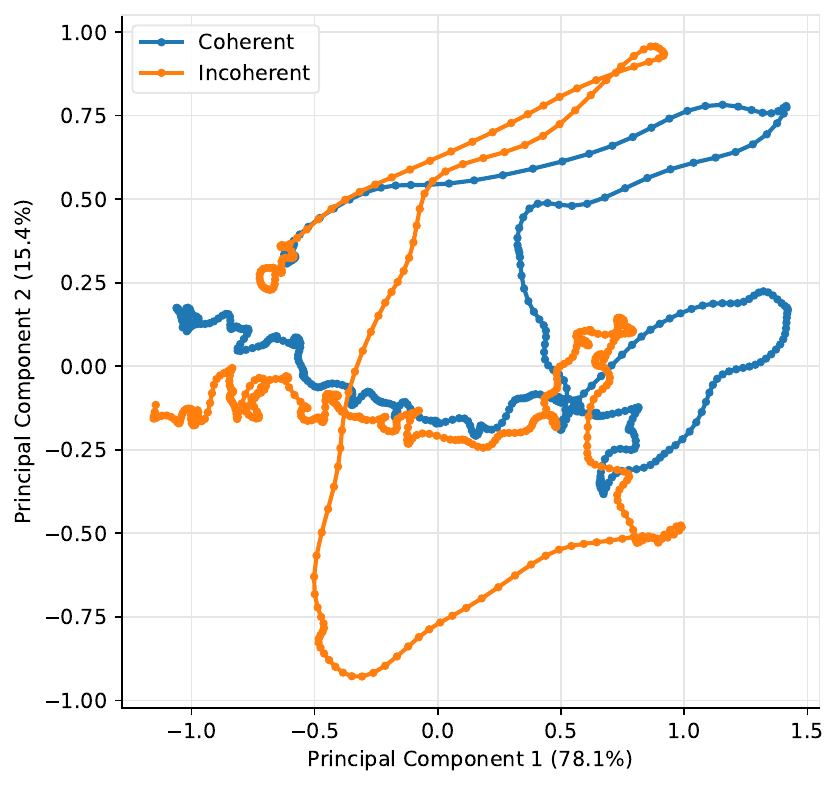}
        \subcaption{\emph{City} principal components.}
        \label{fig:City_PC12_parametric}
     \end{subfigure}
     \caption{Spatial PCA analysis of the scene-matched hand-annotated cut-TRPs.}
    \label{fig:PCA_Spatial}
\end{figure*}

\begin{figure*}[h!]
     \centering
     \begin{subfigure}[b]{.49\textwidth}
        \centering
        \includegraphics[width=\textwidth]{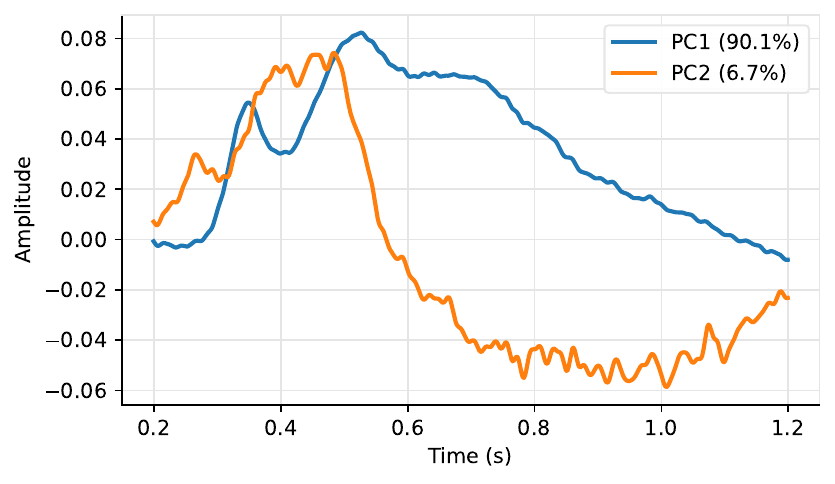}
        \subcaption{\emph{Art} principal eigenvectors.}
        \label{fig:Art_DL_eigenvectors}
     \end{subfigure}
     \hfill
     \begin{subfigure}[b]{.49\textwidth}
        \centering
        \includegraphics[width=\textwidth]{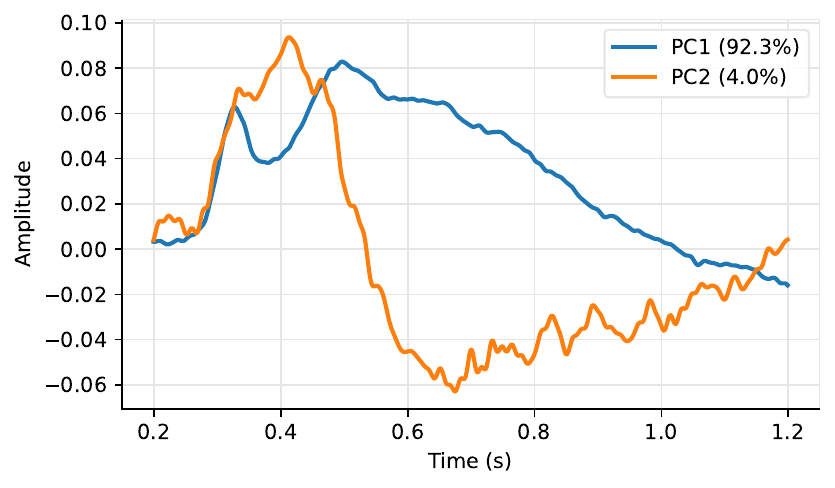}
        \subcaption{\emph{City} principal eigenvectors.}
        \label{fig:City_DL_eigenvectors}
     \end{subfigure}
     \vskip 3mm
     \begin{subfigure}[b]{.242\textwidth}
        \centering
        \includegraphics[width=\textwidth]{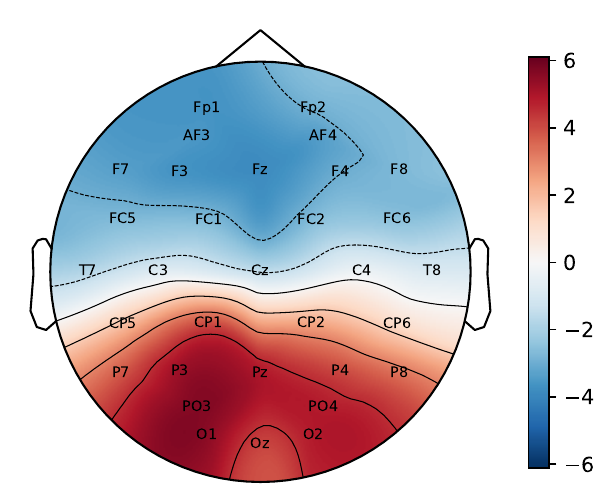}
        \subcaption{\emph{Art} coherent PC1.}
     \end{subfigure}
     \hfill
     \begin{subfigure}[b]{.242\textwidth}
        \centering
        \includegraphics[width=\textwidth]{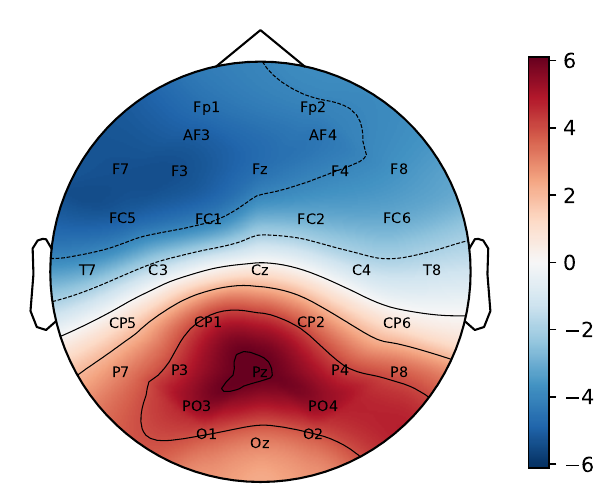}
        \subcaption{\emph{Art} incoherent PC1.}
     \end{subfigure}
     \hfill
     \begin{subfigure}[b]{.242\textwidth}
        \centering
        \includegraphics[width=\textwidth]{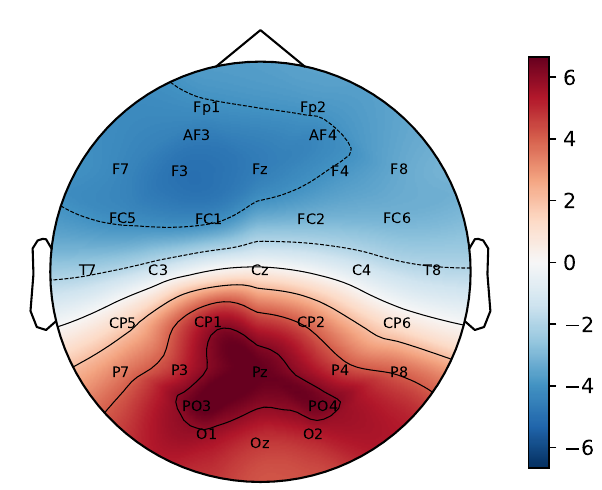}
        \subcaption{\emph{City} coherent PC1.}
     \end{subfigure}
     \hfill
     \begin{subfigure}[b]{.242\textwidth}
        \centering
        \includegraphics[width=\textwidth]{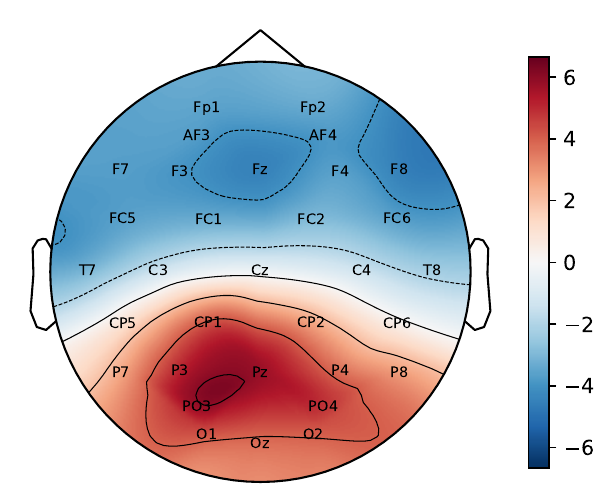}
        \subcaption{\emph{City} incoherent PC1.}
     \end{subfigure}
     \vskip 3mm
     \begin{subfigure}[b]{.242\textwidth}
        \centering
        \includegraphics[width=\textwidth]{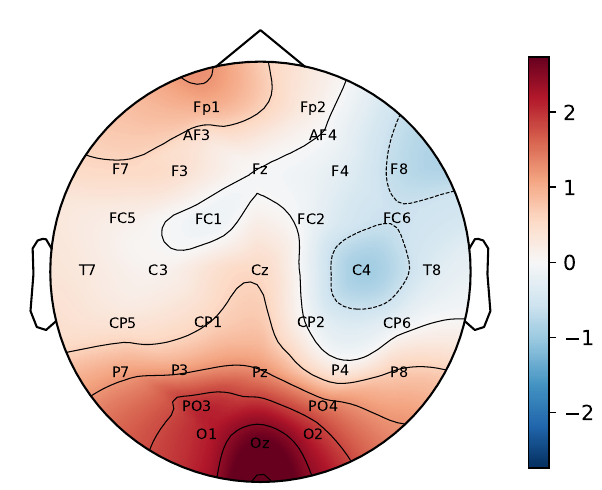}
        \subcaption{\emph{Art} coherent PC2.}
     \end{subfigure}
     \hfill
     \begin{subfigure}[b]{.242\textwidth}
        \centering
        \includegraphics[width=\textwidth]{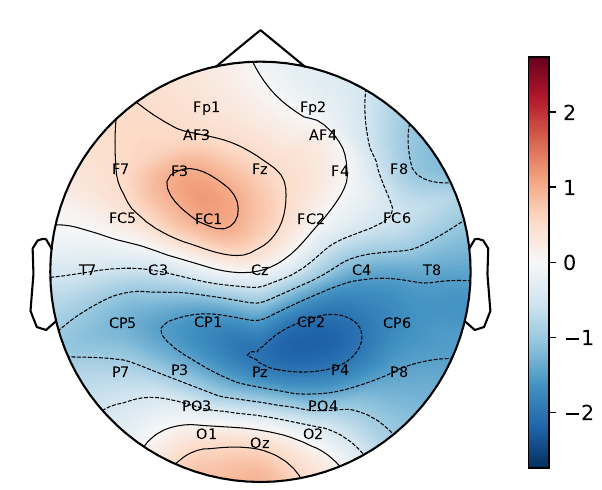}
        \subcaption{\emph{Art} incoherent PC2.}
     \end{subfigure}
     \hfill
     \begin{subfigure}[b]{.242\textwidth}
        \centering
        \includegraphics[width=\textwidth]{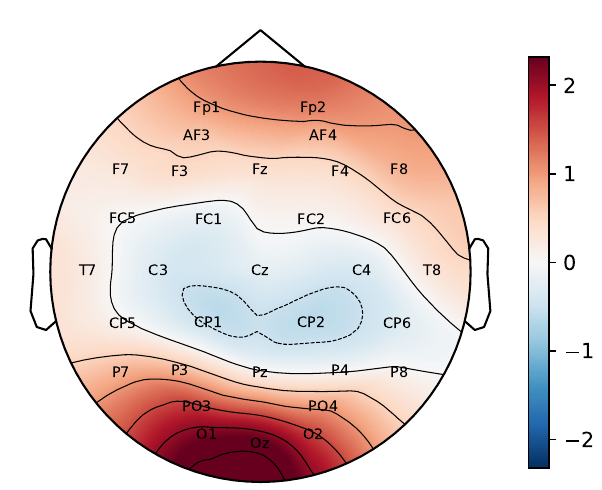}
        \subcaption{\emph{City} coherent PC2.}
     \end{subfigure}
     \hfill
     \begin{subfigure}[b]{.242\textwidth}
        \centering
        \includegraphics[width=\textwidth]{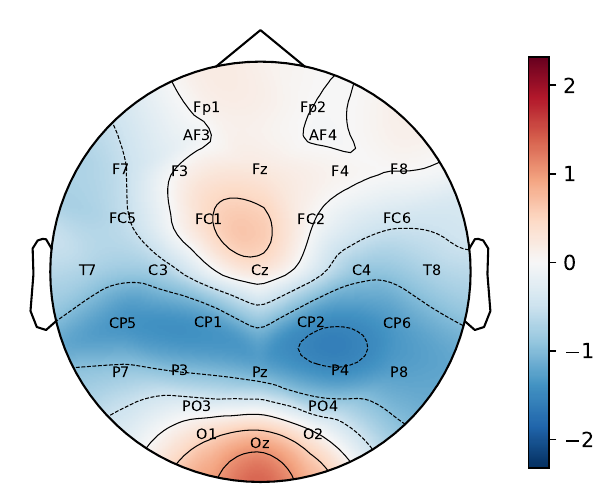}
        \subcaption{\emph{City} incoherent PC2.}
     \end{subfigure}
     \caption{Temporal PCA analysis of the TRPs from the detected cuts.}
    \label{fig:PCA_temporal_DL}
\end{figure*}

\end{document}